\documentclass[letterpaper,11pt]{article}
\usepackage{jheppubMod}
\usepackage{adjustbox}
\usepackage{academicons}      
\usepackage{array,multirow}
\usepackage{bm}  %
\usepackage{booktabs}
\usepackage{float}
\usepackage[T1]{fontenc} 
\usepackage[symbol]{footmisc}
\usepackage{braket,bm}
\usepackage{mathrsfs}
\usepackage{mathtools}
\usepackage{slashed,soul}
\usepackage{tikz}
\usetikzlibrary{arrows,decorations.pathmorphing,backgrounds,positioning,fit,petri,automata,shadows,calendar,mindmap,
decorations.markings,calc}
\usepackage{xspace}
\usepackage{slashed}
\usepackage{array,multirow}
\usepackage{graphicx}
\usepackage{graphics}
\usepackage{epsfig}
\usepackage{amsfonts}
\usepackage{amsmath}
\usepackage{amssymb}
\usepackage{dsfont}
\usepackage{color}
\usepackage{pdflscape}
\usepackage{pifont}
\usepackage{pgfplots}
\usepackage{pgfplotstable}
\usepackage{bbold}
\usepackage[normalem]{ulem}
\usepackage{xcolor,cancel,youngtab}

\usepackage{hyperref}
\hypersetup{
    colorlinks=true,       
    linkcolor=blue,          
    citecolor=blue,        
    filecolor=blue,      
    urlcolor=blue           
}

\definecolor{labelkey}{rgb}{0,0.5,0.0}

\newcommand{\ep}{\epsilon}

\newcommand{\al}{\alpha}
\newcommand{\bt}{\beta}
\newcommand{\g}{\gamma}

\newcommand{\vL}{\ensuremath{\mathcal{L}}}    
    \newcommand{\Dt}{\Delta}
\newcolumntype{P}[1]{>{\centering\arraybackslash}p{#1}}

\newcommand{\GeV}{{\rm ~GeV}}
\newcommand{\TeV}{{\rm ~TeV}}

\newcommand{\fb}{{\rm ~fb}}
\newcommand{\ab}{{\rm ~ab}}
\newcommand{\invfb}{{\rm ~fb^{-1}}}
\newcommand{\invab}{{\rm ~ab^{-1}}}

\newcommand{\ga}{\gamma}

\newcommand{\beq}{\begin{equation}}
\newcommand{\eeq}{\end{equation}}
\newcommand{\be}{\begin{equation}}
\newcommand{\ee}{\end{equation}}
\newcommand{\bea}{\begin{eqnarray}}
\newcommand{\eea}{\end{eqnarray}}
\newcommand{\ben}{\begin{eqnarray*}}
\newcommand{\een}{\end{eqnarray*}}

\newcommand{\bma}{\begin{pmatrix}}
\newcommand{\ema}{\end{pmatrix}}

\renewcommand{\Re}[1]{{{\rm Re\,} #1}}

\def\lixo#1{}

\def\slashchar#1{\setbox0=\hbox{$#1$}           
  \dimen0=\wd0                                    
  \setbox1=\hbox{/} \dimen1=\wd1                  
  \ifdim\dimen0>\dimen1                           
    \rlap{\hbox to \dimen0{\hfil/\hfil}}            
    #1                                             
  \else                                          
    \rlap{\hbox to \dimen1{\hfil$#1$\hfil}}        
    /                                           
 \fi}                                           %

\newcommand{\Or}{\mathcal O}

\newcommand{\vp}{\varphi}
\newcommand{\sq}{^{2}}

\newcommand{\dslash}[1]{#1 \llap{/\kern-0.5pt}}
\newcommand{\Dslash}[1]{#1 \llap{/\kern+1.5pt}}
\newcommand{\DDslash}[1]{#1 \llap{/\kern+2.3pt}}
\newcommand{\dslashh}[1]{#1 \llap{/\kern+1pt}}

\definecolor{cadmiumgreen}{rgb}{0.0, 0.42, 0.24}
\definecolor{darkpastelgreen}{rgb}{0.01, 0.75, 0.24}
\definecolor{darkspringgreen}{rgb}{0.09, 0.45, 0.27}
\definecolor{forestgreen(web)}{rgb}{0.13, 0.55, 0.13}
\definecolor{forestgreen(traditional)}{rgb}{0.0, 0.27, 0.13}
\definecolor{cobalt}{rgb}{0.0, 0.28, 0.67}
\definecolor{darkblue}{rgb}{0.0, 0.0, 0.75}
\definecolor{darkred}{rgb}{0.55, 0.0, 0.0}
\definecolor{palatinatepurple}{rgb}{0.41, 0.16, 0.38}
\definecolor{burntorange}{rgb}{0.8, 0.33, 0.0}

\newcommand{\nn}{\nonumber}

\def\12{\frac{1}{2}}
\def\0nbb{$0\nu\beta\beta$}

\title{Leptonic anomalous magnetic moments in $\nu$SMEFT}

\author[a]{Vincenzo Cirigliano,}
\author[b]{Wouter Dekens,}
\author[c,d]{Jordy de Vries,}
\author[a]{Kaori Fuyuto,} 
\author[a]{Emanuele Mereghetti,}
\author[e]{Richard Ruiz,}

\affiliation[a]{Theoretical Division, Los Alamos National Laboratory, Los Alamos, NM 87545, USA}
\affiliation[b]{Department of Physics, University of California at San Diego, La Jolla, CA 92093, USA}
\affiliation[c]{Institute for Theoretical Physics Amsterdam and Delta Institute for Theoretical Physics,\\ University of Amsterdam, Science Park 904, 1098 XH Amsterdam, The Netherlands}
\affiliation[d]{Nikhef, Theory Group, Science Park 105, 1098 XG, Amsterdam, The Netherlands}
\affiliation[e]{Institute of Nuclear Physics, Polish Academy of Sciences,\\ ul. Radzikowskiego, Cracow 31-342, Poland}

\abstract{
We investigate contributions to the anomalous magnetic moments of charged  leptons in the neutrino-extended Standard Model Effective Field Theory ($\nu$SMEFT). We discuss how $\nu$SMEFT operators can contribute to a lepton's magnetic moment at one- and two-loop order. We show that only one operator can account for existing electronic and muonic discrepancies, assuming new physics appears above $1$ TeV. In particular, we find that a right-handed charged current in combination with minimal sterile-active mixing can explain the discrepancy for sterile neutrino masses of $\mathcal O(100)$ GeV while avoiding direct and indirect
constraints. We discuss how searches for sterile neutrino production at the (HL-)LHC, measurements of $h\rightarrow \mu^+ \mu^-$ and searches for  $h\rightarrow e^+ e^-$, neutrinoless double beta decay experiments, and improved unitarity tests of the CKM matrix can further probe the relevant parameter space. 
\looseness-1
}


\preprint{IFJPAN-IV-2021-6, LA-UR-21-24456}

\arxivnumber{2105.11462}

\begin{document}
\maketitle
\setcounter{page}{2}
\flushbottom

\section{Introduction}
The E989 experiment at Fermilab~\cite{Grange:2015fou}, which has recently measured the anomalous magnetic moment of the muon $a_\mu$ to the highest precision~\cite{Abi:2021gix}, confirms the larger-than-predicted measurement of the E821 experiment at Brookhaven~\cite{Bennett:2006fi} and could
at last indicate the existence of beyond-the-Standard-Model (BSM) physics outside the neutrino and dark sectors. 
A combination of the two measurements gives the average
\begin{equation}
a^{\rm average}_\mu = (116\,592\,061\pm 41) \cdot 10^{-11}.
\end{equation}
This should be compared to the Standard Model (SM) prediction of
\begin{equation}\label{amuSM}
a^{\rm SM\,\,\,\,\,\,\,}_\mu = (116\,591\,810\pm 43) \cdot 10^{-11}\,,
\end{equation}
 as reported in the white paper by the global theory initiative \cite{Aoyama:2020ynm}, 
based on the calculations of Refs.~\cite{Czarnecki:2002nt,Melnikov:2003xd,Aoyama:2012wk,Gnendiger:2013pva,Kurz:2014wya,Colangelo:2014qya,Davier:2017zfy,Masjuan:2017tvw,Colangelo:2017fiz,Hoferichter:2018kwz,Keshavarzi:2018mgv,Colangelo:2018mtw,Hoferichter:2019mqg,Davier:2019can,Aoyama:2019ryr,Keshavarzi:2019abf,Gerardin:2019vio,Bijnens:2019ghy,Colangelo:2019uex,Blum:2019ugy},
and implies a discrepancy of 
\begin{equation}
\Delta a_\mu = (251 \pm 59) \cdot 10^{-11}.
\label{eq:deltaAMMmu}  
\end{equation}
Of course, further experimental and theoretical investigations must carefully examine possible sources of uncertainty. For example:
the recent lattice-QCD calculation of the hadronic vacuum polarization contribution to $\Delta a_\mu$ \cite{Borsanyi:2020mff} 
can shift the SM prediction much closer to the experimental result, essentially eliminating the discrepancy. However, this 
would simultaneously introduce tensions elsewhere \cite{Crivellin:2020zul, Keshavarzi:2020bfy, Colangelo:2020lcg}. Here we will use Eq.~\eqref{amuSM} as our SM prediction for $a^{\rm SM}_\mu$.

The anomalous magnetic moment of the electron is measured with even higher accuracy~\cite{Hanneke_2008}: 
\begin{equation}
 a_e^{\rm exp} = \left(1\, 159\, 652\, 180.73 \pm 0.28 \right) \cdot 10^{-12}.
\end{equation}
In this case, the error of the SM prediction is dominated by the extraction of the fine structure constant $\alpha_{\rm em}$ from experiments that measure the recoil of alkali atoms. The two most recent extractions of $a_e$ from Cesium \cite{Parker_2018} and Rubidium \cite{Morel:2020dww} differ by more than 5$\sigma$, and lead to    
\begin{eqnarray}\label{eq:aeExp}
 \Delta a_e^{\rm Cs} &=& a_e^{\rm exp} - a_e^{\rm SM,\, Cs} = -\left( 0.88 \pm 0.36 \right) \cdot 10^{-12}\,, \nn\\
 \Delta a_e^{\rm Rb} &=& a_e^{\rm exp} - a_e^{\rm SM,\, Rb} = \left( 0.48 \pm 0.30 \right) \cdot 10^{-12}\,.
\end{eqnarray}
This indicates a $-2.4\sigma$ ($+1.6\sigma$) tension with the SM that is not too significant in comparison to the muon case. However, certain new physics models that can explain $\Dt a_\mu$ only predict deviations in $\Dt a_e$ at the $\mathcal{O}(10^{-13})$ level due to an $(m_e/m_\mu)$ suppression~\cite{Giudice:2012ms}. For the $\tau$ lepton, the existing bounds on the anomalous magnetic moment are extracted from measurements of $ e^+e^-\to e^+e^-\tau^+\tau^-$~\cite{Abdallah:2003xd} and 
$e^+e^-\to \tau^+\tau^-$~\cite{GonzalezSprinberg:2000mk}. The respective constraints~\cite{Zyla:2020zbs}
\bea\label{atau}
 - 0.052 < &a_\tau^{e^+e^-\to e^+e^-\tau^+\tau^-} &< 0.013\,, \nn\\
  - 0.007 <& a_\tau^{e^+e^-\to \tau^+\tau^-} &< 0.005\,,
\eea
are still one order of magnitude away from the SM prediction of $a_\tau  = \left(117\, 721 \pm 5 \right) \cdot 10^{-8}$ \cite{Eidelman:2007sb}. While at the moment
$a_\tau$ is not very sensitive to BSM physics, the limit in Eq.\ \eqref{atau} will be improved at Belle-II and at the Large Hadron Collider (LHC)~\cite{Eidelman:2016aih,Chen:2018cxt,Fomin:2018ybj,Beresford:2019gww}. 

The muonic discrepancy in particular has lead to an overwhelming amount of theoretical work looking for suitable explanations. For a recent overview, see Ref.~\cite{Athron:2021iuf}. Most models involve adding new fields and interactions to the SM that go on to generate novel loop contributions to $a_\mu$ and $a_e$. Under the assumption that new fields are heavy\footnote{We note that exceptions exist, for instance in the form of light dark photons \cite{Pospelov:2008zw} or axionlike particles \cite{Marciano:2016yhf}.}, motivated by the absence of evidence for BSM physics in direct searches at the LHC
and elsewhere, these models can be efficiently described by effective field theory (EFT) techniques. A recent investigation 
in the framework of the Standard Model EFT (SMEFT) concluded that only a handful of dimension-six SMEFT operators could account for $\Delta a_\mu$ while not being excluded by other measurements and still obey the EFT demand that $v/\Lambda \ll 1$ \cite{Buttazzo:2020eyl,Aebischer:2021uvt}. Here, $v = \sqrt{2}\langle H^0 \rangle \simeq 246$ GeV is the Higgs field's vacuum expectation value (vev) and $\Lambda$ is the scale of BSM physics. In the EFT language, any BSM scenario that explains $\Delta a_\mu$ will most likely need to generate either the muon dipole operators themselves, 
$\mathcal{O}_{eB},~\mathcal{O}_{eW}$,
which induce corrections to $a_\mu$ at tree level, or semi-leptonic tensor operators, $\mathcal{O}_{lequ}^{(3)}$, which contribute to the dipole moments through one-loop diagrams \cite{Aebischer:2021uvt}. The latter can be significant due to top (or charm) loops, which give rise to a chirally enhanced contributions and scale as $\sim m_{t(c)}/m_\mu$. Similar conclusions hold in case when trying to explain corrections to $a_e$ at the level of Eq.\ \eqref{eq:aeExp}.

In this work, we investigate whether slightly relaxing the SMEFT assumptions could give rise to additional mechanisms. We do so by considering the neutrino-extended SMEFT ($\nu$SMEFT) \cite{delAguila:2008ir,Liao:2016qyd}. This framework is motivated by the need to account for neutrino masses, whose explanation involves right-handed (RH) neutrinos in many (but not all) BSM scenarios. A minimal SM extension would be the addition of a set of $\nu_R$ fields that, in combination with the SM's left-handed (LH) neutrino $\nu_L$ and Higgs fields, can generate neutrino Dirac masses through the usual Higgs mechanism. However, being a gauge singlet, nothing forbids the presence of a RH Majorana mass term $m_{\nu_R}$ for $\nu_R$.
The simultaneous presence of a Dirac and Majorana masses leads to 
three light active Majorana neutrinos and $n$ sterile Majorana states, where $n$ is the number of $\nu_R$ fields that are added to the theory. \textit{A priori} nothing can be said about the RH Majorana masses of sterile neutrinos and they might as well be light, with $m_{\nu_R} \leq v$. While the neutrinos look sterile at low energies, in a large class of BSM models they interact with SM and BSM fields at high energies. Examples are leptoquark models  \cite{Dorsner:2016wpm}, left-right-symmetric models \cite{Pati:1974yy,Mohapatra:1980yp}, 
gauged baryon and lepton number models \cite{Langacker:1980js,Hewett:1988xc,Faraggi:1990ita},
and Grand Unified Theories \cite{Bando:1998ww}. At lower energies, such interactions can be described in terms of  local  operators in the $\nu$SMEFT Lagrangian. Here we consider the effect of $\nu_R$ operators on $a_\mu$ and $a_e$. 

This paper is organized as follows: In Sect.~\ref{nuSMEFT} we introduce the $\nu$SMEFT Lagrangian and its rotation into the mass basis. In Sect.~\ref{loops}, we investigate which $\nu$SMEFT operators can generate corrections to $a_\mu$ and $a_e$ at the one- and two-loop levels and which operators are phenomenologically viable. Sect.~\ref{RHC} focuses on the most promising $\nu$SMEFT operator and discusses its contributions to $ a_{e,\mu}$. We investigate in detail what parameter space can account for $\Delta a_{e,\mu}$ and how this parameter space can be tested in the future in Sect.\ \ref{sec:constraints}. Our main findings are then discussed in Sect.~\ref{eq:results}.
Finally, we conclude in Sect.~\ref{conclusion}.

\section{The $\nu$SMEFT Lagrangian}\label{nuSMEFT} 

We consider an EFT
that consists of the SM supplemented by 
$n$
SM gauge singlets $\nu_R$ (although we mainly consider $n=1$), and include higher-dimensional operators up to dimension six. The Lagrangian of the resulting EFT,  known as $\nu$SMEFT,
can then be written as~\cite{Liao:2016qyd}
\begin{eqnarray}\label{eq:smeft}
	\mathcal L &=&  \mathcal L_{SM} + \bar \nu_R\, i\slashed{\partial}\nu_R- \left[ \frac{1}{2} \bar \nu^c_{R} \,\bar M_R \nu_{R} +\bar L \tilde H Y_\nu \nu_R + \rm{H.c.}\right]\nn \\
	&&+  \mathcal L^{( 5)}_{\nu_L}+  \mathcal L^{(5)}_{\nu_R}+  \mathcal L^{( 6)}_{\nu_L} +  \mathcal L^{( 6)}_{\nu_R} \,,
\end{eqnarray}
where $\mathcal L_{SM}$ contains renormalizable operators of the SM and do not involve $\nu_R$. We use $\Psi^c = C \bar \Psi^T$ for a spinor field $\Psi$ in terms of  $C = - C^{-1} = -C^T = - C^\dagger$, the charge conjugation matrix. $L=(\nu_L,\, e_L)^T$ is the LH
lepton doublet with generation indices suppressed for clarity. $\tilde H = i \tau_2 H^*$ with $H$ denoting the Higgs doublet 
\begin{equation}
	H = \frac{v}{\sqrt{2}} U(x) \left(\begin{array}{c}
		0 \\
		1 + \frac{h(x)}{v}
	\end{array} \right)\,.
\end{equation}
Here, $h(x)$ is the Higgs boson field and $U(x)$ is an $SU(2)$ matrix encoding the electroweak (EW) Goldstone modes. $\nu_{R}$ is a column vector of $n$ RH neutrinos such that $Y_\nu$ is a $3\times n$ Yukawa matrix. $\bar M_R$ is a complex, symmetric $n \times n$ mass matrix containing the Majorana masses of $\nu_R$. The terms $\mathcal L^{( 5)}_{\nu_L}$ and  $\mathcal L^{( 6)}_{\nu_L}$  were constructed in Refs.\ \cite{Weinberg:1979sa} and \cite{Grzadkowski:2010es}, respectively, and describe the dimension-five and -six operators containing just SM fields (no $\nu_R$). Similarly, $\mathcal L^{( 5)}_{\nu_R}$ and  $\mathcal L^{( 6)}_{\nu_R}$ contain the complete set of dimension-five and -six operators involving at least one $\nu_R$ field and were derived in Refs.\ \cite{Aparici:2009fh} and \cite{delAguila:2008ir,Liao:2016qyd}, respectively.

The dimension-five terms generate Majorana masses for
$\nu_L$ and $\nu_R$
and can be written as
\be\label{eq:dim5masses}
 \mathcal L^{(5)}_{\nu_L} = \ep_{kl}\ep_{mn}(L_k^T\, C^{( 5)}\,CL_m )H_l H_n\,,\qquad  \mathcal L^{(5)}_{\nu_R}=- \bar \nu^c_{R} \,\bar M_R^{(5)} \nu_{R} H^\dagger H\,,
\ee
where $C^{(5)}$ and $M_R^{(5)}$ are dimensionful Wilson coefficients.
Throughout this work we employ a notation where Wilson coefficients of gauge-invariant $\nu$SMEFT operators have absorbed all factors of the EFT cutoff scale $\Lambda$. The coefficient of an operator of dimension $d$ then has mass dimension $4-d$ and scales as $C^{(d)}\sim \Lambda^{4-d}$. 

Transition dipole moments appear at the same mass dimension as the terms in Eq.\ \eqref{eq:dim5masses}, but require at least two sterile states.   The first interesting contributions to $\Dt a_\ell$ for $\ell=e,\mu$, involving only one $\nu_R$, then arise from dimension-six operators. The effects of the operators in (normal) SMEFT on the leptonic dipole moments were recently discussed in Ref.\ \cite{Aebischer:2021uvt} and here we focus on the dimension-six operators involving right-handed neutrinos, which are collected in Table \ref{tab1}.

\begin{table}
\centering
\resizebox{\columnwidth}{!}{
\begin{tabular}{|c|c|c|c|c|c|}
\hline
 \multicolumn{2}{|c|}{$\psi^2H^3$} & \multicolumn{2}{|c|}{$\psi^2H^2D$} & \multicolumn{2}{|c|}{$\psi^2HX(+\mbox{H.c.})$}
\\
\hline
$\mathcal{O}_{L\nu H}(+\mbox{H.c.})$ & $(\bar{L}\nu_R )\tilde{H}(H^\dagger H)$ &
$\mathcal{O}_{H\nu }$ &  $(\bar \nu_R \gamma^\mu \nu_R )(H^\dagger i \overleftrightarrow{D_\mu} H)$&
$\mathcal{O}_{\nu B}$ & $(\bar{L}\sigma_{\mu\nu}\nu_R )\tilde{H}B^{\mu\nu}$
\\
& &
$\mathcal{O}_{H\nu e}(+\mbox{H.c.})$ & $(\bar{\nu}_R \gamma^\mu e)({\tilde{H}}^\dagger i D_\mu H)$ &
$\mathcal{O}_{\nu W}$ &$(\bar{L}\sigma_{\mu\nu}\nu_R )\tau^I\tilde{H}W^{I\mu\nu}$
\\
\hline
\multicolumn{2}{|c|}{$(\bar{R}R)(\bar{R}R)$}  &   \multicolumn{2}{|c|}{$(\bar{L}L)(\bar{R}R)$} &   \multicolumn{2}{|c|}{$(\bar{L}R)(\bar{L}R)(+\mbox{H.c.})$}
\\
\hline
$\mathcal{O}_{\nu \nu }$ & $(\bar \nu_R \gamma^\mu \nu_R )(\bar \nu_R \gamma_\mu \nu_R )$ &
$\mathcal{O}_{L\nu }$ & $(\bar{L}\gamma^\mu L)(\bar \nu_R \gamma_\mu \nu_R )$ &
$\mathcal{O}_{L\nu Le}$ & $(\bar{L}\nu_R )\epsilon(\bar{L}e)$
\\
$\mathcal{O}_{e\nu }$ & $(\bar{e}\gamma^\mu e)(\bar \nu_R \gamma_\mu \nu_R )$ &
$\mathcal{O}_{Q\nu }$ & $(\bar{Q}\gamma^\mu Q)(\bar \nu_R \gamma_\mu \nu_R )$ &
$\mathcal{O}_{L\nu Qd}$ & $(\bar{L}\nu_R )\epsilon(\bar{Q}d)$
\\
$\mathcal{O}_{u\nu }$ & $(\bar{u}\gamma^\mu u)(\bar \nu_R \gamma_\mu \nu_R )$ &
& &
$\mathcal{O}_{LdQ\nu }$ & $(\bar{L}d)\epsilon(\bar{Q}\nu_R )$
\\
$\mathcal{O}_{d\nu }$ & $(\bar{d}\gamma^\mu d)(\bar \nu_R \gamma_\mu \nu_R )$&
& &
&
\\
$\mathcal{O}_{du\nu e}(+\mbox{H.c.})$ & $ (\bar{d}\gamma^\mu u)(\bar \nu_R \gamma_\mu e)$&
& &
&\\
\hline
\multicolumn{2}{|c|}{$(\bar{L}R)(\bar{R}L)$}  &
\multicolumn{2}{|c|}{$(\slashed{L}\cap B)(+\mbox{H.c.})$}  &   \multicolumn{2}{|c|}{$(\slashed{L}\cap\slashed{B})(+\mbox{H.c.})$}
\\ \hline
$\mathcal{O}_{Qu\nu L}(+\mbox{H.c.})$ & $(\bar{Q}u)(\bar \nu_R L)$ &
$\mathcal{O}_{\nu \nu \nu \nu }$ & $(\bar \nu_R^c\nu_R )(\bar \nu_R^c \nu_R )$ &
 $\mathcal{O}_{QQd\nu }$ & $\epsilon_{ij}\epsilon_{\alpha\beta\sigma}(Q^i_{\alpha}CQ^j_{\beta})(d_{\sigma}C \nu_R )$
\\
& &
& &
$\mathcal{O}_{udd\nu }$ & $\epsilon_{\alpha\beta\sigma}(u_{\alpha}Cd_{\beta})(d_{\sigma}C\nu_R )$
\\
\hline
\end{tabular}
} 
\caption{The complete basis of dimension-six operators  involving $\nu_R$, taken from Ref.~\cite{Liao:2016qyd}. Operators are expressed in terms of a column vector of $n$ gauge singlet fields, $\nu_R$,
and SM fields, i.e., the lepton and Higgs doublets, $L$ and $H$,  left-handed quark doublet $Q = (u_L, d_L)^{T}$, and right-handed fields $e$, $u$, and $d$.  
}
\label{tab1}
\end{table}

We work in  the basis where charged leptons $e^i_{L,R}$ and quarks $u^i_{L,R}$, $d^i_R$ are mass eigenstates ($i=1,2,3$). After EW symmetry breaking this means that 
$d^{i,\,\rm gauge}_L = V^{ij} d_L^{j,\,\rm mass}$, 
where $V$ is the Cabibbo-Kobayashi-Maskawa (CKM) matrix. At dimension six, 
the mass terms of the neutrinos after EW symmetry breaking but before mass-diagonalization take the form,
  \bea
 \mathcal L_m &=& -\frac{1}{2} \bar {\mathcal N}^c M_\nu \mathcal N +{\rm H.c.}\,,\qquad M_\nu = \bma M_L &M_D^*\\M_D^\dagger&M_R^\dagger \ema \,,\nn\\
M_L &=& -v^2 C^{(5)}\,,\qquad 
M_R = \bar M_R + v^2 \bar M_R^{(5)}\,,\qquad
M_D =\frac{v}{\sqrt{2}} \left[Y_\nu -\frac{v^2}{2}C_{L\nu H}^{(6)}\right]\,.
\eea 
Here $\mathcal N = (\nu_L,\, \nu_R^c)^T$ making $M_\nu$ a $(3+n)\times (3+n)$ symmetric matrix. This matrix can be diagonalized by a  transformation involving the unitary matrix $U$, given by
\bea\label{Mdiag}
U^T M_\nu U =m_\nu = {\rm diag}(m_1,\dots , m_{3+n})\,, \quad\text{with}\quad \mathcal N = U \mathcal N_m\,.
\eea
The  mass eigenstates can be written as $\nu = \mathcal N_m+\mathcal N_m^c=\nu^c$. The flavor eigenstates are then related by $\nu_L = P_L (PU)\nu$ and $\nu_R = P_R (P_sU^*)\nu$, where $P = \begin{pmatrix}\mathcal I_{3\times 3} & 0_{3 \times n}  \end{pmatrix}$
and $P_s = \begin{pmatrix} 0_{n\times 3} & \mathcal I_{n \times n}  \end{pmatrix}$ are projection matrices. For simplicity we will often consider the case $n=1$, implying four massive neutrinos with masses $m_i \ll \Lambda$. The first three states $\nu_{1,2,3}$ approximately correspond to active neutrinos with 
sub-eV masses~\cite{Aker:2019uuj}, while $\nu_4$ is a nearly sterile neutrino with mass $m_4$ that could be significantly larger. For $n=1$, the mass terms involving $\nu_R$, $M_D$, and $M_R$, are not sufficient to generate the masses of the light neutrinos. Two sterile states are in principle sufficient, but would require significant tuning 
to obtain the sizable mixing matrix elements that are needed to resolve  the discrepancy in the lepton's magnetic moments. We will assume a non-zero value of $M_L$ that accounts for the measured neutrino masses and mixing. Such an $M_L$ can for instance be induced at tree level by integrating out even heavier neutrino mass eigenstates and $SU(2)$ triplets, or at loop level via more complicated field content~\cite{Ma:1998dn,Bonnet:2012kz,Cai:2017jrq}.

Throughout this work we generically denote the mixing between active, lepton-flavor states $\alpha\in\{e,\mu,\tau\}$ with (light and heavy) neutrino mass eigenstates $j \in \{1,\dots , 3+n\}$ by $U_{\alpha j}$. From oscillation data, $\vert U_{\alpha j}\vert\sim\mathcal{O}(1)$ for $j=1,2,3$, while from various constraints (as described below) $\vert U_{\alpha j}\vert\ll1$ for $j\geq4$. In this notation, the $3\times3$ Pontecorvo--Maki--Nakagawa--Sakata (PMNS) matrix that parametrizes active-light neutrino mixing is non-unitary and is embedded in the $(3+n)\times(3+n)$ mixing matrix. For the case of $n=1$, we label the active-sterile mixing matrix element associated with $\nu_4$ by $U_{\alpha4}$. In addition, we adopt the mixing formalism of Ref.~\cite{Atre:2009rg} and denote the mixing between ``sterile'' flavors $S = \{1,\dots n\}$ with 
mass eigenstates $j = \{1,\dots , 3+n\}$ by $U_{Sj}$. By unitarity, $\vert U_{Sj}\vert\ll1$ for $j=1,2,3$, while
$\vert U_{Sj}\vert$ can be $\mathcal{O}(1)$ for $j\geq4$.

\begin{table}[t!]\renewcommand{\arraystretch}{1.2}
\centering
\resizebox{\columnwidth}{!}{

\begin{tabular}{|c||c|cc|cccc|}
\hline\hline
class&$\psi^2H^3$ & \multicolumn{2}{|c|}{$\psi^2H^2D$} &  \multicolumn{4}{|c|}{$(\bar{R}R)(\bar{R}R)$}
\\
$\mathcal C_i=v^2 C_i$&$\mathcal  C_{L\nu H}$ & $\mathcal C_{H\nu } $&  \color{forestgreen(web)}{$^*\mathcal C_{H\nu e }$}& $\mathcal C_{\nu \nu}$ & $\mathcal C_{u \nu,d \nu}$& $\mathcal C_{e \nu}$& $\color{forestgreen(web)}{^*\mathcal C_{du \nu e}}$\\
\hline
$b_i$	& $L^2 Y_\nu^\dagger Y_e$ & $L^2 Y_e$ & \color{forestgreen(web)}{$L Y_\nu$ }&  $L^3 Y_e \, Y_\nu^\dagger Y_\nu$& $L^2 Y_e\{Y_\nu Y_\nu^\dagger,L\}$ & $L^2 Y_e$ & $L^2 Y_u^\dagger Y_d Y_\nu$ \\
$\Dt a_\mu/(\mathcal C_i \Dt a_\mu^{{\rm expt.}})$ 	& $10^{-2} Y_\nu$ & $10^{-2} $ & \color{forestgreen(web)}{$3\cdot 10^{3} \,Y_\nu$} &  $10^{-4}Y_\nu^\dagger Y_\nu $ & $10^{-2}\{Y_\nu Y_\nu^\dagger,10^{-2}\}$ & $10^{-2}$ & $0.4 Y_\nu^\dagger$ \\
$\Dt a_e/(\mathcal C_i\Dt a_e^{{\rm expt.}})$ 	& $10^{-3} Y_\nu$ & $10^{-3} $ & \color{forestgreen(web)}{$4\cdot 10^{4} \,Y_\nu$} &  $10^{-5}Y_\nu^\dagger Y_\nu $ & $10^{-3}\{Y_\nu Y_\nu^\dagger,10^{-2}\}$ & $10^{-3}$ & $\color{forestgreen(web)}{6 Y_\nu^\dagger}$ \\
\hline\hline
class& \multicolumn{1}{|c|}{$\psi^2HX(+\mbox{H.c.})$} & \multicolumn{2}{|c|}{$(\bar{L}L)(\bar{R}R)$} & \multicolumn{3}{c}{$(\bar{L}R)(\bar{L}R)$}  &  \multicolumn{1}{|c|}{$(\bar{L}R)(\bar{R}L)$} 
\\
$\mathcal C_i=v^2 C_i$ &  \color{forestgreen(web)}{$^*\mathcal C_{\nu B,\nu W}$} & $\mathcal C_{Q\nu}$ & $\mathcal C_{L\nu}$ &\color{forestgreen(web)}{$^*\mathcal  C_{L\nu Le}$ }& $\mathcal C_{L\nu Q d } $& $\mathcal C_{Ld Q\nu}$ & \multicolumn{1}{|c|}{$\mathcal C_{ Qu\nu L}$}\\
\hline
{$b_i$} 	& \color{forestgreen(web)}{$L Y_\nu^\dagger Y_e$} & $L^2 Y_e\{Y_\nu Y_\nu^\dagger,L\}$& $L^2 Y_e$&  \color{forestgreen(web)}{$L^2 Y_\nu^\dagger$} & $L^2 Y_d^\dagger\,Y_\nu^\dagger Y_e$ & $L^2 Y_d^\dagger\,Y_\nu^\dagger Y_e$ & \multicolumn{1}{|c|}{$L^2 Y_u^\dagger\,Y_e^\dagger Y_\nu $}\\
$\Dt a_\mu/(\mathcal C_i\Dt a_\mu^{{\rm expt.}})$ & \color{forestgreen(web)}{$2 Y_\nu^\dagger$ }& $10^{-2}\{Y_\nu Y_\nu^\dagger,10^{-2}\}$& $10^{-2}$ &\color{forestgreen(web)}{$20 Y_\nu^\dagger$} & $10^{-4} Y_\nu$& $10^{-4} Y_\nu^\dagger $& \multicolumn{1}{|c|}{$10^{-2} Y_\nu$}\\
$\Dt a_e/(\mathcal C_i\Dt a_e^{{\rm expt.}})$ & $0.1 Y_\nu^\dagger$ & $10^{-3}\{Y_\nu Y_\nu^\dagger,10^{-2}\}$& $10^{-3}$ &\color{forestgreen(web)}{$300 Y_\nu^\dagger$} & $10^{-5} Y_\nu$& $10^{-5} Y_\nu^\dagger $& \multicolumn{1}{|c|}{$10^{-3} Y_\nu$}\\
\hline\hline
\end{tabular}
} 
\caption{Estimates of the contributions of the dimension-six Wilson coefficients
to the coefficients  $b_i$ (defined in Eq.~\eqref{eq:ai}),
where $\Delta a_{\ell}=(4m_{\ell}/v) b_i{\cal C}_i$. The third (fourth) row is the ratio $\Delta a_{\ell}/\Delta a_{\ell}^{\rm expt.}$
for $\ell=\mu~(e)$ in units of $\mathcal C_i=v^2 C_i$. 
 $L$ stands for a loop factor and $Y_i$ are the different Yukawa couplings. 
The cases with ${c}_i,~{c}_i^{\prime} \geq 1$ are highlighted (in green) with an asterisk.} 
\label{tab:est}
\end{table}

\section{Contributions to $a_\ell$ from dimension-six $\nu_R$ operators}\label{loops}
In principle, several of the dimension-six operators involving sterile neutrinos will give rise to loop-level corrections to $\Dt a_e$ and $\Dt a_\mu$, which we henceforth denote as $\Dt a_\ell$. In the EFT below the EW scale, 
$\Dt a_\ell$ is generated at tree-level for lepton $e$ by the leptonic dipole moment
\beq
\vL_{\rm dipole} = L_{\substack{e\ga\\ pr}}  \,\bar e_{L\, p} \sigma^{\mu\nu}e_{R\, r} \,F_{\mu\nu }+{\rm H.c.}\,,
\eeq 
where $p,r$ are flavor indices, and the Wilson coefficient $L_{\substack{e\ga \\ \ell\ell}}$ gives a contribution to
$a_\ell$ of
\bea
\Dt a_\ell = \frac{4 m_\ell}{e} {\rm Re}\left(L_{\substack{e\ga \\ \ell\ell}}\right)\,. \eea
In the denominator is the quantity $e=|e|=\sqrt{4\pi\alpha_{\rm em}}$.
 
We now write the loop-level contributions of 
$\nu$SMEFT
interactions to $L_{\substack{e\ga \\ \mu\mu}}$ and $L_{\substack{e\ga \\ e e}}$ as
\begin{equation}
L_{\substack{e\ga \\ \ell \ell}}
=  \left[ e \frac{b_i}{v} (v^2 C_i) \right]_{\ell \ell}\,.
\label{eq:ai}
\end{equation}
{The dimensionless factors $b_i$ parametrize the contributions of gauge and Yukawa couplings as well as the loop order at which the corresponding $\nu$SMEFT operator $\mathcal{O}_i$ generates the dipole interactions.

We do not pursue a comprehensive calculation of $b_i$, and hence $L_{\substack{e\ga \\ \ell \ell}}$, for all operators in Table \ref{tab1},
as simple order-of-magnitude estimates show that most interactions cannot induce sufficiently large anomalous magnetic moments to account for measured  $\Dt a_\mu$. We provide such estimates for all $B$- and $L$-conserving dimension-six $\nu$SMEFT operators  in Table \ref{tab:est}, where we show a na\"ive expression for each generated $L_{e\ga}$,
 as well as its effect on $\Dt a_\mu$ and $\Dt a_e$,
which can be written as
\bea
\Dt a_\ell = \frac{4 m_\ell}{e} {\rm Re}\left(L_{\substack{e\ga \\ \ell\ell}}\right)
\, = \frac{4 m_\ell}{v} {\rm Re}\left({b_i} v^2 C_i\right)\,.
\eea
For the numerical estimates we use $L = (4\pi)^{-2}$ for loop factors and $Y_{e,d,u}=\sqrt{2}m_{\mu(e),b,t}/v$ for $\Delta a_{\mu}~(\Delta a_{e})$. As we assume $\vert\mathcal  C_i\vert  = v^2 \vert C_i\vert$ and $\vert Y_\nu\vert$ to be $\ll \Or(1)$, contributions of the form  $\Dt a_\ell/\Dt a_\ell^{\rm expt.}= c_i \mathcal C_i$ or $\Dt a_\ell/\Dt a_\ell^{\rm expt.}= c_i' Y_\nu \mathcal C_i$, with $c_i,c^{\prime}_i\leq \Or(1)$, are unlikely to explain $\Delta a_\mu^{\rm expt.}$ or $\Delta a_e^{\rm expt.}$. In only five cases do we estimate that ${c}_i,~{c}_i^{\prime} \geq 1$ and denote these with asterisks.

In many cases the impact on $\Dt a_\ell$ is too small because the first possible 
diagrammatic insertions appear beyond one-loop order, or 
because of the appearance of small Yukawa couplings.
For example: in the operator class $(\overline{L}R)(\overline{R}L)$, the impact of $\mathcal{O}_{Qu\nu L}$ on $\Dt a_{\mu}$ is  suppressed by $b_{Qu\nu L}\sim L^2 Y_u^\dagger Y_e^\dagger Y_\nu\sim 10^{-8}Y_\nu$, which leads to the ratio $\Dt a_\mu/\Dt a_\mu^{\rm expt.} \sim  10^{-2} \times Y_\nu  v^2 C_{Qu\nu L}$.
Only in the extreme limit where $ Y_\nu  \sim v^2 C_{Qu\nu L} \sim \mathcal{O}(10)$ and the EFT breaks down
can $\mathcal{O}_{Qu\nu L}$ account for experimental measurements.
In summary, while the contributions from most operators are significantly smaller than 
the current discrepancies, 
five
operators generate potentially interesting contributions. Indicated (in green) with asterisks in Table \ref{tab:est}, these 
operators are:
\begin{align}
 \mathcal{O}_{H\nu e}, \quad 
 \mathcal{O}_{\nu B}, \quad 
 \mathcal{O}_{\nu W}, \quad
  \mathcal{O}_{L\nu Le} \quad \text{and}\quad 
 \mathcal{O}_{du\nu e}.
 \label{eq:opList}
\end{align}
The corresponding one- and two-loop contributions to $\Delta a_\ell$ are depicted in Fig.~\ref{dim6_diagram}. The left diagram is a threshold contribution induced by the 
$\mathcal{O}_{H\nu e}$
operator.  
The two diagrams in the middle depict the dipoles, $\mathcal{O}_{\nu B}$ and $\mathcal{O}_{\nu W}$, and four-lepton operators, $ \mathcal{O}_{L\nu Le}$, which generate magnetic moments through renormalization group running. In the electron case, the semileptonic operator $\mathcal O_{du\nu e}$ also induces a somewhat sizable corrections to $\Delta a_e$ as in the right diagram.

\begin{figure}
\center 
\includegraphics[width=1\textwidth]{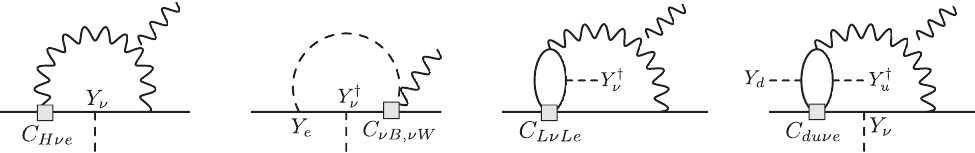}
\caption{Examples of the contribution from the dim-6 operators to $\Delta a_{l}$. The gray box represents an insertion of a $\nu$SMEFT operator.  The solid line corresponds to fermions, while dashed and wavy lines denote the $SU(2)$ Higgs doublet and a SM gauge boson, respectively. }
\label{dim6_diagram}
\end{figure}

\subsection{Dimension-six dipole and four-fermion interactions}
We first discuss the dipole operators, $\mathcal{O}_{\nu B}$, $\mathcal{O}_{\nu W}$, which could, in principle, give significant contributions to $ a_\mu$. These couplings also generate magnetic moments $\mu_\nu$ for the active neutrinos that are described at tree level by the interaction
$\vL\supset-\frac{1}{4}[\mu_{\nu}]_{ij}{\nu}^T_iC\sigma^{\mu\nu}P_R\nu_jF_{\mu\nu}+ {\rm h.c.}$ \cite{Bell:2005kz,Grimus:2000tq},
where $C$ is the charge conjugation matrix and $i,j$ run over light neutrino mass eigenstates\footnote{The coefficient $\mu_\nu$ used here corresponds to $\lambda^\dagger$ defined in Ref.\ \cite{Grimus:2000tq}.}.
Barring cancellations, current experiment limits  require $v^2\big|U^*_{\ell i}\left[C_{\nu B,\nu W}\right]_{\ell S} U^*_{Sj}\big|\lesssim 10^{-6}$~\cite{Giunti:2014ixa,Zyla:2020zbs}.
Since  $\vert U_{\ell i}\vert\sim\mathcal{O}(1)$, the Wilson coefficients $C_{\nu B}$ and $C_{\nu W}$ can only be sizable if the sterile-light mixing element $U_{Sj}$ saturates the constraint with $\vert U^*_{Sj} \vert \sim \mathcal{O}(10^{-6})$. (Smaller values of  $U_{Sj}$ require that $v^2 C_{\nu B,\nu W} \gtrsim \mathcal{O}(1)$, implying a breakdown of the EFT.) Notably, this value of $\vert U_{Sj} \vert$ is consistent with the high-scale Type I seesaw featuring  $\nu_4$ with mass $m_4\sim\mathcal{O}(100)$ GeV, i.e.,
$\vert U_{Sj} \vert \sim \vert U_{\ell 4} \vert \sim \sqrt{m_{j}/m_4} \sim 10^{-6}$.
Nevertheless, this limit in conjunction with Table~\ref{tab:est} implies that there is no room for significant contributions to either lepton's magnetic moment. 

The third promising operator that could be relevant for both $a_\mu$ and $a_e$ is $\mathcal{O}_{L\nu L e}$. While this four-fermion interaction contributes to active neutrinos' magnetic moments, the contributions are suppressed by a small lepton mass in addition to a loop factor. Hence, no strong constraint can be set. We therefore go beyond the na\"ive estimate of Table \ref{tab:est} and first consider its leading-log contributions to the coefficient $L_{\substack{e\ga \\ \mu\mu}}$. The required RGEs are those of the mixing of $C_{L\nu Le}$ into the leptonic dipole moments $C_{eB}$ and $C_{eW}$, as well as the running and mixing among $C_{eB}$ and $C_{eW}$. The former are given by 
\bea
\frac{d}{d\ln\mu} C_{\substack{ eB\\pr}} = \frac{1}{2}\frac{g_1 {\rm y}_\ell}{16\pi^2}C_{\substack{L\nu Le \\ pStr}} \left[Y^*_\nu\right]_{tS}\,, \qquad \frac{d}{d\ln\mu} C_{\substack{ eW\\pr}} = -\frac{1}{4}\frac{g}{16\pi^2}C_{\substack{L\nu Le \\ pStr}} \left[Y^*_\nu\right]_{tS}\,,
\eea
where ${\rm y}_\ell = -1/2$  is the hypercharge of charged lepton $\ell$,
the indices $p,r,t =\{ e,\,\mu,\, \tau\}$ indicate the charged lepton flavor, and $S = \{1,\dots n\}$ label sterile neutrino flavors. This induces the combination of dipole moments that couples to the $Z$ boson, but not the photon, below the EW
scale. Thus, to generate the electromagnetic dipole moment one requires the mixing between $C_{eB}$ and $C_{eW}$ as well. Using the RGEs of Refs.\ \cite{Jenkins:2013zja,Jenkins:2013wua} we obtain
\bea
\Dt a_\mu/\Dt a_\mu^{{\rm expt.}}& \simeq &0.2\,v^2 C_{\substack{L\nu Le \\ \mu St\mu }} \left[Y^*_\nu\right]_{tS} \ln^2\left (\frac{m_4}{\Lambda}\right)\,,\nn\\
\Dt a_e/\Dt a_e^{{\rm Cs}}& \simeq &-2.8\,v^2 C_{\substack{L\nu Le \\ e St e }} \left[Y^*_\nu\right]_{tS} \ln^2\left (\frac{m_4}{\Lambda}\right)\,.
\eea

As estimate, setting $\nu_4$'s mass to $m_4\sim 100$ GeV and $\Lambda = 1$ TeV gives coefficients of $\Or(1) \times v^2 C_{\substack{L\nu Le \\ \mu St\mu }}$ and $\Or(10) \times v^2 C_{\substack{L\nu Le \\ e Ste }}$, which are both a factor $20$ smaller than the na\"ive estimates in Table \ref{tab:est}.
The quantities $v^2 C_i$ and the relevant entry of the Yukawa coupling $\left[Y_\nu\right]_{\ell S}\sim U_{\ell i} m_i U_{S i}/v$ are constrained to be well below $\Or(1)$, either due to the EFT assumptions requiring
$m_4 < \Lambda$ or experimental constraints. In particular,  the mixing elements are bounded by $\vert U_{e4}\vert \lesssim 6\cdot 10^{-2}$ and $\vert U_{\mu4}\vert \lesssim 10^{-2}$ (see 
discussions below), making significant contributions unlikely in the range of validity of the EFT, $\Lambda\gg v$. For the same reasons we exclude the, in principle, promising contributions to $a_e$ from $\Or_{du\nu e}$. We conclude that only 
$\mathcal{O}_{H\nu e}$
provides a contribution that can be large enough to explain the discrepancies surrounding $a_e$ and $a_\mu$.

\subsection{A dimension-six right-handed current}\label{RHC}
We now focus on the remaining dim-6 operator $O_{H\nu e}$, which in the gauge basis is given by
\bea
\vL_{H\nu e}^{
} &=&    \left(i \tilde{\vp}^{\dagger} D_{\mu} \vp\right) \, \bar \nu_R \gamma^\mu \,C_{H\nu e} e_R
+  \mathrm{H.c.}\, ,
\label{dim6edms}
\eea
where $D_\mu = \partial_\mu -ig_1 {\rm y} B_\mu -i\frac{g}{2} \tau^I  W_\mu^I-ig_s t^a G^a_\mu$ is the usual SM covariant derivative, and $C_{H\nu e}$ is generally an $n\times 3$ matrix in flavor space with mass dimension GeV$^{-2}$. In the unitary gauge after  EW symmetry breaking, this operator can be written in the neutrino mass basis as
\bea\label{RH}
\vL_{H\nu e}^{
}& =& \frac{g v^2}{2\sqrt 2}\sum_{j=1}^{n+3} \,\bar \nu_j \g^\mu  \left[U^TP_s^T C_{H\nu e}\right]_{j \alpha} e_{R\alpha}\,W_\mu^+  \left(1+\frac{h}{v}\right)^2 +\text{H.c.}
\eea

Focusing on couplings between charged leptons and  heavy neutrino states,  we obtain
\bea\label{RH2}
\vL_{H\nu e}& \approx & \frac{g v^2}{2\sqrt 2} \sum_{j=4}^{n+3} \left[\bar C_{H\nu e}\right]_{j \alpha} \, ( \bar \nu_j \g^\mu \, e_{R\alpha})\,W_\mu^+  \left(1+\frac{h}{v}\right)^2 +\text{H.c.},
\eea
where $\alpha \in \{e, \mu,\tau\}$, $S$ again denotes the sterile flavor states, and we have defined the effective coupling $\left[\bar C_{H\nu e} \right]_{j \alpha}\equiv U_{Sj}(C_{H\nu e})_{S \alpha }$. For $n$ heavy mass eigenstates,  this interaction between sterile neutrinos and RH charged leptons  leads to the following one-loop correction to $\Delta a_\ell$: 
\bea\label{DeltaamuRH}
\Delta a_\ell  &=& - \frac{2m_\ell}{(4\pi)^2}  \sum_{j=4}^{n+3} m_j\,w(x_j)\,{\rm Re}\left( U_{\ell j}   \left[\bar  C_{H\nu e}\right]_{j \ell}\right) \,. 
\eea
Here $x_j = m_j^2/m_W^2$, and the loop function $w(x)$ is given by 
\bea
w(x) = \frac{4+x(x-11)}{2(x-1)^2}+3\frac{x\sq \ln x}{(x-1)^3}\, .
\label{eq:loopFn}
\eea

As expected from the earlier estimate, this contribution to a lepton's anomalous magnetic moment is chirally enhanced by the sterile neutrinos' masses. The same loops induce contributions to leptonic electric dipole moments (EDMs) proportional to ${\rm Im}[ U_{\ell j}   \left[\bar  C_{H\nu e}\right]_{j \ell}]$. These contributions to $d_\ell$ can be obtained by multiplying Eq.\ \eqref{DeltaamuRH} by the factor $\left[-e/(2m_\ell)\right]$ and replacing Re$\left[\dots \right]\to$Im$\left[\dots\right]$. Although the experimental limit on the electron EDM \cite{Andreev:2018ayy} leads to stringent constraints on the CP-violating phase of the couplings with $\ell=e$, it probes a different combination of couplings than $\Dt a_\ell$. In what follows we consider the most important experimental probes of the sterile couplings in Eq.\ \eqref{DeltaamuRH}, namely  $\bar C_{H\nu e}$ and $U_{\ell4}$.

\section{Experimental probes of the leptonic right-handed current}\label{sec:constraints}

We now discuss experimental constrains and projected sensitivity at near-future facilities for the RH current operator $\mathcal{O}_{H\nu e}$ and the active-sterile mixing element $U_{\ell4}$. Discussions on the parameter space preferred by $\Dt a_\mu$ and $\Dt a_e$ are deferred to Section~\ref{eq:results}.

\subsection{Contributions to $h\to \ell\ell$}\label{sec:hll}
Following Ref. \cite{Alioli:2017ces}, we consider the contributions of 
$\mathcal{O}_{H\nu e}$
to the Yukawa couplings of the charged leptons. Modifications of these Yukawa couplings are a common signature of explanations for the anomalies involving magnetic moments~\cite{Fajfer:2021cxa,Crivellin:2021rbq}.  In particular, a nonzero $C_{H\nu e}$  can induce large corrections to the lepton's Yukawa coupling $y_\ell $ that are proportional to the sterile neutrino mass $m_j$. We start from
\bea
\vL \supset -m_\ell \, \bar \ell_L \ell_R-y_\ell \, \bar \ell_L \ell_R\, h+{\rm H.c.}\,\, ,
\eea
where $y_\ell$ here is related to the $Y_e$ in Table~\ref{tab:est} by $y_\ell=\sqrt{1/2}\left[ Y_e^\dagger\right]_{\ell\ell}$.
The running of the lepton's Yukawa and mass are modified at a renormalization scale $\mu_r$
by the sterile neutrino as follows: 
\bea \label{YbRGE}
\frac{d m_\ell}{d \ln\mu_r} &=& \frac{1}{2} \frac{g\sq v\sq}{(4\pi)\sq} U_{\ell j} \,\left[\bar C_{H\nu e}\right]_{j\ell}m_j \,(x_j-3)\,,
\\
\frac{d y_\ell}{d \ln\mu_r} &=&  \frac{1}{2}\frac{g\sq v}{(4\pi)\sq}  U_{\ell j} \,\left[\bar C_{H\nu e}\right]_{j\ell} m_j\,(x_h+3x_j-9)\,,
\eea
where $x_{h(j)}=m_{h(j)}\sq/m_W\sq$ and the equations holds for $m_{j}, m_{W}<\mu_r$. These RGEs can be obtained from the contributions of the SMEFT interaction $C_{Hud}$ to the SM down-type Yukawa couplings $Y_d$ and dimension-six operator $C_{dH}$ determined in Refs.\ \cite{Jenkins:2013zja,Jenkins:2013wua,Alonso:2013hga}. Explicitly, one replaces the couplings of Refs.~\cite{Jenkins:2013zja,Jenkins:2013wua,Alonso:2013hga} by
\begin{equation}
 C_{Hud}\to C_{H\nu e}, \quad Y_u \to Y_\nu^\dagger, \quad C_{dH} \to C_{e H}, \quad C_{uH} \to C_{L\nu H}, \quad\text{and}\quad Y_d \to Y_e\,.
 \end{equation}
With this we can make the identification
  \begin{align} 
  y_\ell = \sqrt{1/2}& \left[ Y_e^\dagger-\frac{3}{2} v^2C_{eH}\right]_{\ell\ell}, \quad
  m_\ell = \frac{v}{\sqrt{2}}\left[ Y_e^\dagger-\frac{1}{2} v^2C_{eH}\right]_{\ell\ell}\,,   
    \\
& \text{and} \quad 
 \left[Y_\nu\right]_{pS} = \frac{\sqrt{2}}{v}\left[M_D\right]_{pS} = \frac{\sqrt{2}}{v} U_{pi}m_i U_{Si} \,.
  \end{align}
In addition, threshold corrections to the muon mass and Yukawa appear after integrating out a sterile neutrino or the $W$ boson.

For the lepton mass below the EW scale, we obtain the $\mu_r$-independent result
 \begin{equation}\label{eq:mmu}
m_\ell=m_\ell(\mu_r) -  \frac{m_W^2}{(4\pi)\sq}   \left[ x_j-1+(x_j-3) \log \frac{\mu_r^2}{m_W^2}   -  \frac{x_j -4}{x_j-1}x_j \log x_j \right] U_{\ell j} \,\left[\bar C_{H\nu e}\right]_{j\ell}m_j\,.
 \end{equation}
 The effective Yukawa coupling that is probed in Higgs decays is then given by
 \bea
y_\ell^{\rm (eff)}& =& y_\ell(\mu_r) -\frac{1}{4} \frac{g^2v}{(4\pi)^2} U_{\ell j} \,\left[\bar C_{H\nu e}\right]_{j\ell}m_j \bigg\{( x_h + 3 x_j-9) \log \frac{\mu_r^2}{m_W^2}  - 3\frac{x_j (x_j  -3)}{x_j -1} \log x_j \nn \\
& & +  x_j \beta_j \log\left(\frac{\beta_j -1}{\beta_j + 1} \right) +  (x_h - 2)  \beta_W \log\left(\frac{\beta_W -1}{\beta_W + 1}\right) +  [2 x_h + x_j (x_j - 7)] f_1(x_h,x_j) \nn \\
& & - [4 + (2 - x_h) x_j] f_2(x_h,x_j) + (-5 + 2 x_h + 4 x_j)
\bigg\}\,.
\eea
In the above, the function  $y_\ell(\mu_r)$ is given by the solution of Eq.\ \eqref{YbRGE}, $\beta_{W(j)} = \sqrt{1 - 4 m_{W (j)}^2/m_h^2}$ is a kinematic factor, and the loop functions $f_{1,2}$ are
\begin{eqnarray}
f_{1}(x_h,x_j) = \int_0^1 dz \frac{1}{1 - x_j + x_h z } \log\left(\frac{x_j - x_h (1 - z) z}{1 + (x_j -1) z}\right)\, , \nn \\
f_{2}(x_h,x_j) = \int_0^1 dz \frac{1}{-1 + x_j + x_h z } \log\left(\frac{1 - x_h (1 - z) z}{x_j  + z (1-x_j)}\right)\, .
\end{eqnarray}
We stress that $y_\ell^{\rm (eff)}$ is independent of the renormalization scale $\mu_r$. 

To obtain constraints from measurements and searches for Higgs decays to leptons, we assume only $C_{H\nu e}$ is generated at a scale $\mu_r=\Lambda$. We thus use the measured value of the lepton mass as input for $m_\ell$ in Eq.\ \eqref{eq:mmu}. This determines $m_\ell(\Lambda)$ from which we obtain $y_\ell(\Lambda ) = m_\ell(\Lambda)/v$ and subsequently $y_\ell(\mu_r)$. This procedure sets the BSM Yukawa interactions to zero at $\mu_r=\Lambda$, i.e.\ $C_{eH}(\Lambda)=0$. One might expect a general BSM scenarios to generate multiple operators in the EFT, which could give additional contributions to $h\to\ell\ell$. Our assumptions lead to conservative  constraints as long as there are no significant cancellations between the different contributions. 

For $\ell=\mu$, we compare the obtained value for $|y_\mu^{\rm (eff)}/y_\mu^{\rm (SM)}|^2$ to recent CMS measurements, which using $\mathcal{L}\approx 137\invfb$ of data at $\sqrt{s}=13\TeV$ reports a signal strength  of~\cite{Sirunyan:2020two} 
\begin{equation}
 \mathcal \mu_{h\to\mu\mu} = \frac{{\rm Br}(h\to  \mu^+   \mu^-)_{\phantom{\rm SM}}}{{\rm Br}(h\to  \mu^+   \mu^-)_{\rm SM}}= 1.19^{+0.44}_{-0.42}\,,
 \end{equation} 
and the projected HL-LHC sensitivity of $\mu_{h\to\mu\mu}= 1.0\pm0.1$ \cite{Cepeda:2019klc}. ATLAS has also performed a measurement and reports a comparable result but with a slightly larger uncertainty \cite{Aad:2020xfq}.

We use an analogous procedure to obtain the effective Yukawa coupling of the electron, $y_e^{\rm (eff)}$, which we constrain by using the recent limit Br$(h\to e^+e^-)< 3.6\cdot 10^{-4}$  \cite{Aad:2019ojw}. This translates to
\bea
\left| \frac{y_e^{\rm (eff)}}{y_e^{\rm (SM)}} \right|\leq \bigg\{ \begin{matrix*}[l] 268 & \quad{\rm LHC\, (current)}& \mathcal{L}=  139\, {\rm fb}^{-1} \\ 58 & \quad{\rm HL-LHC}& \mathcal{L}= 3\, {\rm ab}^{-1}\end{matrix*}\,,
\eea
where we used a factor of $\sqrt{139\,{\rm fb}^{-1}/ 3\,{\rm ab}^{-1}}$ to estimate the sensitivity of the HL-LHC from the current limit.
For the reach of a future Higgs factory to BR$(H\to\ell^+\ell^-)$, see Ref.\ \cite{Altmannshofer:2015qra}.

\subsection{Constraints on electromagnetic  interactions of light neutrinos} 
Here, we briefly consider the impact of dimension-six $\nu$SMEFT operators on the magnetic moment of light, active neutrinos. Specifically, a diagram with the same topology as the first diagram in Fig.~\ref{dim6_diagram}, but with neutrino mass eigenstates $\nu_i$ and $\nu_j$ on the external legs and charged leptons $\ell_\alpha$ running inside the loop, generates a photon dipole operator with coefficient 
\begin{equation}
\left[\mu_\nu\right]_{ij} = -2\sqrt{2}U_{\al i}^*\left[ c_w C_{\nu B} + s_w C_{\nu W}\right]_{\al S}U_{Sj}^* = - \frac{e}{(4\pi)^2} \frac{\sqrt{2} m_{\ell_\alpha}}{v} U^*_{Sj} \left[C^{*}_{H\nu e}\right]_{S \alpha} U^*_{\alpha i}.
\end{equation}
For $\nu_j \to \nu_i \gamma$, this corresponds
to a neutrino magnetic moment of
 \begin{equation}\label{munu}
 \left[\mu_\nu \right]_{i j} = \frac{8}{(4\pi)^2} {y_{ \ell_\alpha}} y_e  \mu_B\, U^*_{S j} \left[v^2 C^{*}_{H\nu e}\right]_{S \alpha} U^*_{\alpha i}
\overset{\tiny\al = \mu}{\approx}
  4.4 \cdot 10^{-11} \mu_B  \, U^*_{S j} \left[v^2 C^{*}_{H\nu e}\right]_{S \mu} U^*_{\mu i},
 \end{equation}
 where for the numerical estimate we specified $\alpha=\mu$.
Presently, the Borexino experiment sets an upper bound  of about $\mu_\nu^{(\rm eff)}<2.8\cdot10^{-11}\mu_B$ at 90\% CL on the effective magnetic moment of light neutrinos~\cite{Borexino:2017fbd}. This 
corresponds to limits at the same order of magnitude on the individual elements $\left[\mu_\nu\right]_{ij}$.
 Thus, current limits 
 only probe large coefficients with  $v^2 \vert C^{}_{H\nu e}\vert \sim \mathcal O(1)$ and mixing  angles $\vert U_{S j}\vert \sim \mathcal O(1)$, which
are weak in comparison to the experiments discussed below.

\subsection{Lepton-flavor violation}

The same diagrams that induce $\Delta a_\ell$ also contribute to  lepton-flavor-changing dipoles, and hence $\ell_\beta\to\ell_\alpha \gamma$ transitions,  through the coupling
\bea\label{}
L_{\substack{e\ga\\ \al \bt}}  &=& - \frac{e}{32\pi^2}  \sum_{j=4}^{n+3} U_{\al j} m_j  \left[\bar  C_{H\nu e}\right]_{j \beta} \,w(x_j)\,. 
\eea 
The loop function $w(x)$ is defined in Eq.~\eqref{eq:loopFn}. The branching rate for $\mu \rightarrow e \gamma$ is then given by
\begin{eqnarray}\label{eq:mue}
& & \textrm{BR}(\mu \rightarrow e \gamma) = \tau_\mu \frac{m_\mu^3}{4\pi} \left( \left|L_{\substack{e\ga\\ \mu e}}\right|^2 + \left| L_{\substack{e\ga\\ e \mu}} \right|^2 \right) \nonumber \\
 & &=
 \frac{3 \alpha_{\rm em}}{8 \pi \hat \Gamma_\mu} 
\left( \left|\sum_{j=4}^{n} U_{\mu j} \frac{m_j}{m_\mu}  \left[v^2 \bar  C_{H\nu e}\right]_{j e} \,w(x_j)\right|^2
+ \left|\sum_{j=4}^{n} U_{e j} \frac{m_j}{m_\mu}  \left[v^2 \bar  C_{H\nu e}\right]_{j \mu} \,w(x_j)\right|^2 \right)\,,
  \end{eqnarray}
where we expressed the muon lifetime $\tau_\mu$ as
\begin{equation}
 \tau^{-1}_\mu = \frac{G_F^2 m_\mu^5}{192 \pi^3} \hat \Gamma_\mu.
\end{equation}
The dimensionless number $\hat \Gamma_\mu$ accounts for radiative corrections and finite $m_e$ and is adjusted to reproduce the measured muon lifetime. Using Ref.~\cite{Zyla:2020zbs}, we obtain $\hat \Gamma_\mu = 0.996$. Eq.\ \eqref{eq:mue} should then be compared with the stringent limit on the $\mu \rightarrow e \gamma$ branching ratio,  $\textrm{BR}(\mu \rightarrow e \gamma) < 4.2 \cdot 10^{-13}$ \cite{TheMEG:2016wtm}.

\subsection{Constraints on $U_{\mu 4}$ and $U_{e 4}$ from CKM unitarity} 
From the expression for $\Dt a_\ell$ in  Eq.~\eqref{DeltaamuRH}, we observe that the correction to a lepton's anomalous magnetic moment by a heavy neutrino $\nu_j$ $(4 \leq j \leq n+3)$ through  $\mathcal{O}_{H\nu e}$ is proportional to the combination $(U_{\ell j} \,\left[\bar C_{H\nu e}\right]_{j \ell})$. Constraints on the elements of the mixing matrix $U_{\ell j}$
between active and sterile states have been studied extensively in the literature. These papers mainly focus on minimal scenarios in the absence of higher-dimensional operators but many of the results can be applied here as well. 
For recent reviews, see Refs.~\cite{Fernandez-Martinez:2016lgt,Deppisch:2015qwa,Cai:2017mow,Dentler:2018sju,Bolton:2019pcu}.

For simplicity we consider $n=1$ and begin with discussing the $U_{\mu 4}$ mixing element. Laboratory constraints on $U_{\mu 4}$ are very dependent on the mass of the sterile neutrino. At  small masses $m_4 \lesssim \mathcal O(\rm eV)$, the bounds mainly come from oscillation experiments and require $\vert U_{\mu 4}\vert \lesssim 0.1$. Strong bounds exists for $ \mathcal O(1\,\mathrm{MeV}) < m_4 < m_W$ 
from a large class of  accelerator experiments,  where sterile neutrinos are produced directly in meson and weak boson decays; for details see Sect.~\ref{sec:lhcDirect}.
These bounds weaken significantly for $m_4 > m_W$. To account for the measured value of $\Delta a_\mu$, a larger $m_4$ is preferred to maximize the chiral enhancement. This automatically selects the region $m_W < m_4 < \Lambda$. In this mass range the constraint on $U_{\mu4}$ is driven by muonic, mesonic, and nuclear $\beta$ decays. In particular, the interplay of these limits manifests as stringent constraint on  $U_{\mu 4}$ from CKM unitarity. 

Most of the discussion for $U_{e4}$ is similar to that of the muonic mixing element, with the exception that for light sterile masses, i.e., $m_4 < 1$ MeV. There, a strong bound arise from looking for kinks in the electron spectrum in nuclear $\beta$-decays \cite{Shrock:1980vy,Bryman:2019bjg}. For instance: $\vert U_{e 4}\vert \lesssim 0.03$ for $m_4 \sim 0.1$ MeV from the $\beta$-decay spectrum of ${}^{35}$S \cite{Holzschuh:2000nj}. Because of these constraints and the chiral enhancement, again sterile neutrino masses are preferred above the $W$ boson mass.

We proceed by following Refs.~\cite{Fernandez-Martinez:2016lgt, Blennow:2016jkn} and update some of the theoretical input for the extraction of $V_{ud}$ from superallowed $0^+ \rightarrow 0^+$ decays, and of $V_{us}$ from leptonic and semileptonic kaon decays.
The existence of a sterile neutrino that is not kinematically accessible in muon decay causes a reduction of the muon decay width (relative to SM expectations) due to
the non-unitarity of the
PMNS matrix.  This affects the extraction of the Fermi constant from muon decay through the relationship
\begin{equation}
 G^{(\mu)}_F = 
 G^{(0)}_F \left(1 - \frac{1}{2}|U_{\mu 4}|^2 -\frac{1}{2} |U_{e 4}|^2\right)\,.
\end{equation}
Here,  $G_F^{(0)}=1/\sqrt{2}v^2$ is the Fermi constant in the SM  at tree level, and
we assume $\vert U_{\mu 4}\vert, \vert U_{e 4}\vert \ll 1$. The modification of $G_F$ propagates to the semileptonic decays that are used to extract the $V_{ud}$ and $V_{us}$ 
elements of the  CKM matrix and test first-row unitarity.
In particular,
\begin{eqnarray}
 V^{0^+ \rightarrow 0^+}_{ud} &=&  V^{}_{ud} \left(1 + \frac{1}{2} |U_{\mu 4}|^2 
 \right) \,, \label{Vud1}\\
| V_{us} 
|^{K \to \pi e \nu} &= &  | V_{us} 
|\left(1 + \frac{1}{2} |U_{\mu 4}|^2  \label{Vus1}
 \right)\,, \\
 | V_{us} 
 |^{K \to \pi \mu \nu} &= &  | V_{us} 
  |\left(1 + \frac{1}{2} |U_{e 4}|^2 
 \right)\,.
 \label{Vus2}
 \end{eqnarray}
The branching ratio $\Gamma(K \rightarrow \mu \nu)/\Gamma(\pi \rightarrow \mu \nu)$, which determines $V_{us}/V_{ud}$, is not affected by $U_{\mu 4}$ and $U_{e4}$. 
For $V^{0^+ \rightarrow 0^+}_{ud}$ we use the recent extraction reported in the Particle Data Group (PDG)~\cite{Zyla:2020zbs} ,
which adopts the improved evaluation of radiative corrections from Refs.~\cite{Seng:2018yzq,Seng:2018qru,Czarnecki:2019mwq}.  
We also consider the more conservative analysis of Towner and Hardy (TH)~\cite{Hardy:2020qwl}, 
 which quotes a larger nuclear theory error. Respectively, the extracted values of $V_{ud}^{0^+\to0^+}$ are 
\begin{equation}
\vert V^{0^+\to0^+}_{ud}\vert \big|_{\rm PDG} = 0.97370(14),\qquad 
\vert V^{0^+\to0^+}_{ud}\vert \big|_{\rm TH} = 0.97373(31)\,.
\end{equation}

For the extraction of $V_{us}$, we use the experimental values of $|V_{us} f^{K \pi}(0)|^{K \rightarrow \pi \ell \nu} $ given in Ref. \cite{Moulson:2017ive}, which lists the electron and muon channels separately. The theoretical input consists of the ratio of decay constants $f_K/f_\pi$ and of the form factor $f^{K\pi}(0)$, for which we take the  FLAG`19 average with $N_f =2 + 1 +1$ flavors\cite{Aoki:2019cca}. Assuming unitarity of the CKM matrix, we then perform a two-parameter fit to $\vert U_{\mu 4}\vert $ and $\vert V_{us}\vert = \lambda\simeq \sqrt{1-V_{ud}^2}$.  Since the current theoretical and experimental input indicate $|V_{ud}|^2 + |V_{us}|^2 < 1$ at the $3\sigma$-$4\sigma$ level and the presence of sterile neutrinos would  pull in the opposite direction, i.e., drive $|V_{ud}|^2 + |V_{us}|^2 > 1$, a scenario with $\nu_4$  and $\vert U_{\mu 4}\vert \neq 0$ is disfavored with respect to the SM. This tension could be accounted for by including 
dimension-six SMEFT operators, see e.g.\ \cite{Crivellin:2021njn,Crivellin:2020ebi,Crivellin:2020lzu}. Using  $ V^{0^+\to0^+}_{ud}\big|_{\rm PDG}$, we find 
\begin{eqnarray}\label{Umu4ckm}
 |U_{\mu 4}| < 0.01\,  \quad\text{at}\quad 95\%\, {\rm C.L.}\,,
\end{eqnarray}
with the bound weakening  
to  $|U_{\mu 4}| < 0.02$ if we use $ V^{0^+\to0^+}_{ud}\big|_{\rm TH}$.
The electron case 
is slightly less 
constrained and is less affected by the extraction of $V_{ud}$. Using  $V^{0^+\to0^+}_{ud}\big|_{\rm PDG}$ we find
\begin{eqnarray}\label{Ue4ckm}
 |U_{e 4}| < 0.06\,,  \quad\text{at}\quad 95\%\, {\rm C.L.}\,,
\end{eqnarray}
which becomes  $|U_{e 4}| < 0.07$ with $V^{0^+\to0^+}_{ud}\big|_{\rm TH}$. 

So far we have neglected  $\Or_{H\nu e}$. The contributions of this operator to muonic and mesonic $\beta$ decays scale as $\sim U_{S j}^2 \bar C^2_{H\nu e}$, with $j = 1,2,3$.
Since $\vert U_{S j}\vert \sim \vert U_{\mu 4}\vert\ll1$, these corrections can be safely neglected.

Throughout this discussion, we worked in the context of $n=1$ sterile neutrinos.
Eqs. \eqref{Vud1}-\eqref{Vus2} can be generalized to the $n>1$ case by the replacement  $|U_{\alpha 4}|^2 \rightarrow \sum_{j = 4}^{3+n}|U_{\alpha j}|^2$. The limits in Eqs. \eqref{Umu4ckm} and \eqref{Ue4ckm} then still apply to each $|U_{\alpha j}|$, after marginalizing over other elements.

\subsection{Neutrinoless double beta decay}\label{sec:0nubb}

The impact of sterile neutrinos with non-minimal interactions on neutrinoless double beta decays ($0\nu\beta\beta$)  was studied in Refs. \cite{Cirigliano:2018yza, Dekens:2020ttz}. These constraints are only relevant for interactions with electrons as there is not enough energy available in $0\nu\beta\beta$ to produce final-state muons. Assuming $m_{j} \gtrsim 1$ GeV for $j\geq4$, the heavier neutrinos can be integrated out as $0\nu\bt\bt$ takes place at energies of $\Or({\rm MeV})$. This generates two dimension-nine scalar operators of the form 
\begin{equation}\label{eq:dim9}
 \mathcal L^{(9)} = \frac{1}{v^5} C^{(9)}_{1R}\, \bar q_L \gamma^\mu \tau^+ q_L \, \bar q_L \gamma_\mu \tau^+ q_L \, \bar e_R  e_R^c
 + \frac{1}{v^5} C^{(9)}_{1L}\, \bar q_L \gamma^\mu \tau^+ q_L \, \bar q_L \gamma_\mu \tau^+ q_L \, \bar e_L  e_L^c\,,
\end{equation}
where $q = (u,\, d)^T$,  $\tau^+ = (\tau_1+i\tau_2)/2$, and $C^{(9)}_i$ are dimensionful coefficients given by
\begin{eqnarray}
C^{(9)}_{1\, L}  &=& -2 vV_{ud}^2   \sum^{n+3}_{j=4} \frac{U^2_{ej}}{m_j}    \,,\qquad 
C^{(9)}_{1\, R}  = -\frac{v^5}{2}   V_{ud}^2   \sum^{n+3}_{j=4}\left[\bar C_{H \nu e}\right]^*_{j e} \frac{1}{m_j}   \left[\bar C_{H \nu e}\right]^*_{j e} .
\end{eqnarray}
The second effect of heavy sterile neutrinos is that in the definition of the effective electron neutrino Majorana mass $m_{\beta\beta}$, the sum is restricted to the three lightest neutrinos
\begin{equation}
 m_{\beta\beta}  = \sum_{j=1}^3 m_j U^2_{e j}\,.
\end{equation}
This is the usual expression in SMEFT at dimension five but with the slight difference being that 
the  $3\times 3$ 
PMNS matrix is not by itself unitary. 

The $0\nu\beta\beta$ half-life $T^{0\nu}_{1/2}$ can then be expressed as 
\begin{equation}
\left( T^{0\nu}_{1/2}\right)^{-1} = g_A^4 \Big\{  G_{01} \left( |\mathcal A_L|^2 + |\mathcal A_R|^2 \right)  - 2 (G_{01} - G_{04})\textrm{Re} (\mathcal A_L^* \mathcal A_R)\Big\}\,,
\end{equation}
where $G_{01}$ and $G_{04}$ are phase space factors, and $\mathcal A_L$ and $\mathcal A_R$ are the product of nuclear matrix elements (NMEs),  hadronic couplings and Wilson coefficients of $\nu$SMEFT operators. In our case
\begin{eqnarray}
 \mathcal A_L &=& - \frac{m_{\beta \beta}}{m_e} {V_{ud}^2} \left( - \frac{g_V^2}{g_A^2}
 M_F + M_{GT} + M_T  +\frac{2}{g_A^2} g_\nu^{NN} m_\pi^2 M_{F\, sd}
 \right) + \frac{m_\pi^2}{m_e v} C^{(9)}_{1\, L} \mathcal A_{sd} \,,\nn\\
 \mathcal A_R &=& \frac{m_\pi^2}{m_e v} C^{(9)}_{1\, R} \mathcal A_{sd} \,,\nn\\
 \mathcal A_{sd} &=& 
   \left( \frac{5}{6} g_1^{\pi\pi}
 \mathcal M_{PS, sd} + \frac{1}{2} \left( g_1^{\pi N} - \frac{5}{6} g_1^{\pi\pi} \right)  (M^{AP}_{GT, \, sd} +
 M^{AP}_{T, \, sd}  ) - \frac{2}{g_A^2} g_1^{NN} M_{F,sd}
 \right)\,. 
\end{eqnarray}
Here $M_F$, $M_{GT}$ and $M_T$ are the NMEs induced by the exchange of light Majorana neutrinos, while the `short distance' ($sd$) nuclear matrix elements $\mathcal M_{PS, sd}$, $M^{AP}_{GT, \, sd}$, $M^{AP}_{T, \, sd}$ and  $M^{}_{F, \, sd}$ are the pion- and short-range NMEs induced by the dimension-nine operators $O^{(9)}_{1\,L}$ and $O^{(9)}_{1\,R}$ in Eq.\ \eqref{eq:dim9}, and from the exchange of light neutrinos with hard momenta \cite{Cirigliano:2018hja, Cirigliano:2019vdj}.  For both space phase factors and NMEs, we use the definitions of Ref.\ \cite{Dekens:2020ttz}. 

The low-energy constants (LECs) $g_1^{\pi\pi}$,
$g_1^{\pi N}$ and $g_1^{NN}$ are all expected to be $\mathcal O(1)$.
$g_1^{\pi\pi}$ has been computed on the lattice, yielding
$g_1^{\pi\pi} =0.36 \pm 0.02$ \cite{Nicholson:2018mwc}, while $g_1^{\pi N}$ and $g_{1}^{NN}$ are at the moment unknown. 
We set them to the factorization-inspired values  $g_{1}^{\pi N}=1$ and $g_{1}^{NN}(2~{\rm GeV}) = (1+3g^2_A)/4+1$~\cite{Dekens:2020ttz}.  The short-distance LEC $g_\nu^{NN}$ arises from the exchange of hard Majorana neutrinos and was recently estimated in Refs.~\cite{Cirigliano:2020dmx,Cirigliano:2021qko}. Following Ref.~\cite{Dekens:2020ttz},  for simplicity we set it to $g_\nu^{NN}=-1/(2F_{\pi})^2$ with $F_{\pi}=92.2~$MeV, and note that we will not use the 
amplitude proportional to $m_{\beta \beta}$ in obtaining constraints (see discussion below). 

To keep the expressions compact we now focus on the case of $n=1$. Using the NMEs of Ref.\ \cite{Hyvarinen:2015bda} and the phase space factors extracted from Ref.\ \cite{Horoi:2017gmj}, the half-life of $^{136}$Xe becomes
\begin{eqnarray}
\label{eq:0nubb}
& & \left[ T^{0\nu}_{1/2}\left(^{136}{\rm Xe}\right)\right]^{-1} = \Bigg\{ 2.4 \cdot 10^{-24} \left|\frac{m_{\beta\beta}}{1\, {\rm eV}}\right|^2 \\
&&+ 1.5 \cdot 10^{-7} 
\frac{ \textrm{Re}( m_{\beta\beta} U^{*2}_{e4})}{m_4}
- 7.4 \cdot 10^{-9} \frac{ \textrm{Re}( m_{\beta\beta} \left[v^2 \bar C_{H \nu e}\right]^2_{4 e})}{m_4}   \nonumber\\
& & + 10^{-10} \left(\frac{1\, \rm GeV}{m_4}\right)^2 \left( 23 \,  |U_{e4}|^4 
- 2.3  \, \textrm{Re}\left(\left[v^2 \bar C_{H \nu e}\right]^2_{4 e} U^{\, 2}_{e4}\right) + 1.4  \left|\left[v^2 \bar C_{H \nu e}\right]_{4 e}\right|^4   \right) \Bigg\} \, {\rm yr}^{-1}. \qquad
\nn
\end{eqnarray}
The strongest constraint on the $0\nu\beta\beta$ decay half-life comes from the 2016 KamLAND-Zen measurement \cite{KamLAND-Zen:2016pfg}, which implies the lower limit $T^{0\nu}_{1/2}({}^{136} \mathrm{Xe})>1.07\cdot 10^{26}$ yr at 90\% C.L. Next-generation experiments such as nEXO \cite{Anton:2019wmi} will provide sensitivity at the level of $10^{28}$ yr.

Some care is needed when using these expressions to set limits, as the contributions from $m_{\beta\beta}$ and from $C^{(9)}_{1\, L}$ can depend strongly on the mechanism that accounts for the active neutrino masses. In our present set-up, assuming $M_R\gg M_D\gg M_L$, the light neutrino masses consists of
Weinberg operator and Type I seesaw contributions, i.e., $ M_{\nu_{\rm light}}\simeq M_L -\left[M_DM_R^{-1}M_D^T\right]^\dagger$, which can be generated in a high-scale Type I+II seesaw scenario.
Here, $\left[M_L\right]_{\al\bt} = U_{\al j}^*m_j U_{\bt j}^*$ and we would expect a similar scaling, $\sim U_{\al j}^*m_j U_{\bt j}^*$, for the Type I seesaw contribution. As we consider $m_j> m_W$ and $\vert U_{\al j}\vert\sim 0.01$ for $j\geq4$, a na\"ive estimate of these contributions to the light neutrino masses turns out to be too large $\gtrsim \Or(10)$ MeV. This implies that a realistic scenario of neutrino masses will require non-trivial cancellations, which can also affect the $C_{1\, L}^{(9)}$ contributions. One scenario with such cancellations is that of a low-scale Type I seesaw, e.g., the so-called Inverse Type I seesaw, which features pseudo-Dirac neutrinos \cite{Mohapatra:1986aw,Mohapatra:1986bd,Bernabeu:1987gr}. In such a scenario, the case of $n=2$ corresponds to nearly degenerate $m_4$ and $m_5$. In this limit, we can reproduce the masses and mixing of the light neutrinos with $M_L=0$, and obtain $m_{4,5}>m_W$ simultaneously with $\vert U_{\al j}\vert\sim 0.01$. Apart from giving masses to light neutrinos that are smaller than expected from the above estimate, one also finds that $C_{1\, L}^{(9)}$ scales as $m_{\nu_{\rm light}}/m_4^2$, significantly below the na\"ive estimate $\sim U_{e4}^2/m_4$.
As we do not wish to go into the details of the mass mechanism of neutrinos, we will take a conservative approach and only consider the contributions from $C^{(9)}_{1\, R}$, given by the last term in Eq.\ \eqref{eq:0nubb} when setting limits. This implies that 
$0\nu\beta\beta$ experiments constrain $\left[\bar C_{H \nu e}\right]_{4 e}$ independently of the $U_{4e}$ mixing angle.

The same matching procedure can be followed in the muonic sector, where operators analogous to $C^{(9)}_{1\,L}$
and $C^{(9)}_{1\,R}$ contribute to lepton-number-violating decays such as $K^+ \rightarrow \pi^- \mu^+ \mu^+$, studied in Ref. \cite{Liao:2019gex}.
The limit on the branching ratio, BR$(K^+ \rightarrow \pi^- \mu^+ \mu^+) < 4.2  \cdot 10^{-11}$ \cite{Zyla:2020zbs} is, however, too weak to provide competitive constraints on $[\bar C_{H\nu e}]_{4 \mu}$.

\subsection{Accelerator Constraints and LHC Prospects}
\label{sec:lhcDirect}

If the $\nu$SMEFT operator can indeed account for the measured anomalous magnetic moments of the muon and/or electron, then one anticipates an impact on the production of $\nu_4$ directly through (anomalous) EW currents at accelerator facilities. 
Consequentially, such production is strongly constrained by direct searches. For instance, short baseline and beam dump experiments require $\vert U_{\mu 4}\vert \lesssim 10^{-5}-10^{-3}$ for 150 MeV$\lesssim m_4 \lesssim 450$ MeV~\cite{CortinaGil:2017mqf,Abe:2019kgx}. For larger masses but still below $m_W$, ATLAS~\cite{Aad:2019kiz}, CMS~\cite{Sirunyan:2018mtv,Sirunyan:2018xiv}, and LHCb~\cite{Aaij:2020ovh}  measurements imply that $ \vert U_{\mu 4}\vert \lesssim 3 \cdot 10^{-3}$~\cite{Sirunyan:2018mtv, Aad:2019kiz}. As discussed below, these bounds weaken significantly for $m_4 > m_W$. In this region the strongest indirect limits arise from EW precision data; the strongest direct limits are set by CMS~\cite{Sirunyan:2018mtv,Sirunyan:2018xiv}.

In addition to these experimental limits are theoretical constraints. For ultra heavy $\nu_4$ with $m_4\gg m_W$, the total width of $\nu_4$ $(\Gamma_4)$ grows according to the Goldstone Equivalence Theorem. The perturbativity bound of $\Gamma_4 < m_4$ then requires~\cite{Pascoli:2018heg}:
\begin{equation}
 \vert U_{\mu 4}\vert^2 < \frac{16\pi}{g^2} \frac{m_W^2}{m_4^2}
 \approx 76 \cdot \left(\frac{100\GeV}{m_4}\right)^2.
\end{equation}
As these limits, much like those from $s$-wave unitarity in $W^\pm W^\pm \to \ell^\pm_\alpha \ell^\pm_\beta$ scattering~\cite{Dicus:1991fk,Fuks:2020att}, are much weaker than experimental bounds, we neglect them for the remainder of this study.

\begin{figure}
\center 
\includegraphics[width=\textwidth]{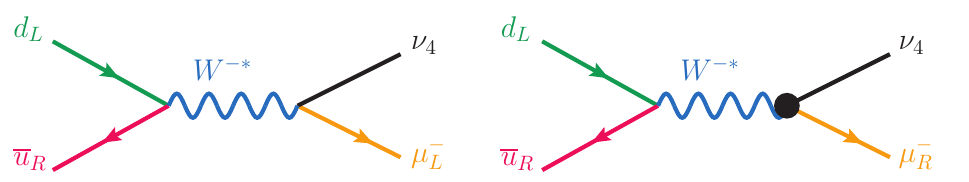}
\caption{
Lowest order, Born graph depicting $d \overline{u}\to W^{-*} \to \mu^-\nu_4$ scattering with leading helicity configurations (subscripts) in $\nu$SMEFT at dimension $d=4$  (L) and  $d=6$ (R). Drawn using \texttt{JaxoDraw}~\cite{Binosi:2008ig}.
} 
\label{fig:diagram_CCDY}
\end{figure}

We now consider the sensitivity of ATLAS and CMS to direct production of $\ell\nu_4$ pairs through dimension-four and -six interactions. Such projections and constraints of dimension-six $\nu$SMEFT operators have been previously estimated in Refs.~\cite{Duarte:2016caz,Ruiz:2017nip,Cai:2017mow,Cottin:2021lzz} and references therein. We mainly focus here on the muonic case and discuss couplings to electrons at the end of this section. For the sterile neutrino masses under consideration, i.e., $m_4 \sim 100-1000\GeV$, the production of $\mu^\pm\nu_4$ pairs due to minimal mixing ($d=4$ operators) is dominated by two processes. The first is the charged-current Drell-Yan (DY) process~\cite{Keung:1983uu}
\begin{equation}
 q \overline{q} \to W^\pm \to  \mu^\pm \nu_4\,,
\end{equation}
as shown in the left diagram of Fig.~\ref{fig:diagram_CCDY} for the $d_L\overline{u}_R\to \mu^-_L \nu_4$ partonic configuration. An additional contribution (not shown) comes from $W\gamma$ fusion~\cite{Datta:1993nm,Dev:2013wba,Alva:2014gxa}, and proceeds through
\begin{equation}
 q \gamma \to  \mu^\pm \nu_4 q'.
\end{equation}
Both processes have matrix elements that are linear in active-sterile mixing $U_{\mu4}$ and therefore exhibit production-level cross sections $\sigma(pp \to \mu^\pm\nu_4+X)$ that scale as $\vert U_{\mu4}\vert^2$.

At the dimension-six level ($d=6$) both channels receive corrections from  $\mathcal{O}_{H\nu e}$. This is shown diagrammatically in the right panel of Fig.~\ref{fig:diagram_CCDY}, for the $d_L\overline{u}_R\to \mu^-_R\nu_4$ partonic channel. As stressed above, the $d=4$ contribution occurs through a LH chiral coupling while the $d=6$ contribution occurs through a RH chiral coupling. Since the muon can be treated as massless in typical LHC collision, the  DY and $W\gamma$ mechanisms at $d=4~(6)$ are driven almost exclusively by LH~(RH) $\mu^-$. (In the same manner, the RH~(LH) $\mu^+$ drives the charge-conjugated processes.)

The helicity amplitudes for RH~(LH) $\mu^-$ at $d=4~(6)$ are helicity-inverting and hence result in squared amplitudes that are suppressed by a factor of $\vert \mathcal{M}_{\rm helicity-inverting}\vert^2\propto m_\mu^2/Q^2 \sim m_\mu^2/m_4^2$. Here $Q$ is the invariant mass of the $(\mu\nu_4)$-system and scales as $Q\sim m_4$. For the range of $m_4$ that is of interest, this translates to a helicity suppression of about $m_\mu^2/m_4^2 \lesssim 10^{-6}$. Similarly, interference between $d=4$ and $d=6$ requires projecting RH~(LH) chiral operators on LH~(RH) helicity states of $\mu^-$, and again is suppressed by a factor of $\Re[\mathcal{M}^*_{d=4}\mathcal{M}_{d=6}] \propto m_\mu^2 / Q^2 \sim m_\mu^2 / m_4^2$.

\begin{figure}
\center 
\includegraphics[width=.47\textwidth]{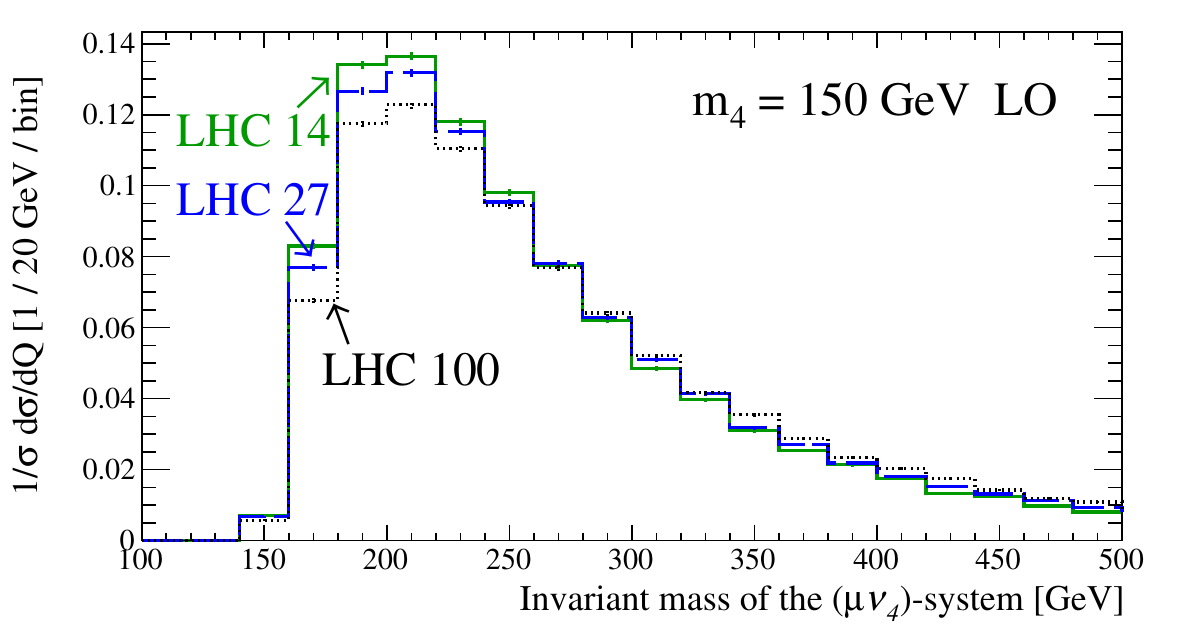}
\includegraphics[width=.47\textwidth]{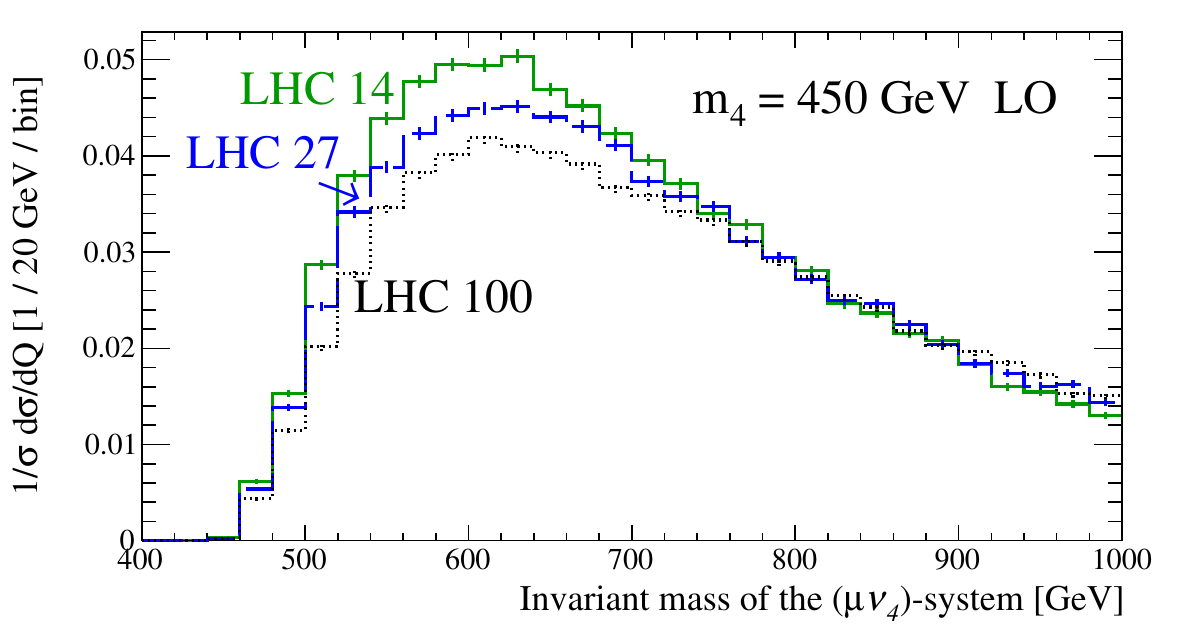}
\caption{The normalized LO invariant mass distribution of the $(\mu\nu_4)$-system, when mediated by the DY process at $d=4$, for (L) $m_4=150\GeV$ and (R) $m_4=450\GeV$, for $\sqrt{s}=14$ (solid), 27 (dash), and 100 (dot) TeV. Adapted from Ref~\cite{Pascoli:2018heg}.} 
\label{fig:issVeto_DYX_LHCMulti_XLO_Mmuv4}
\end{figure}

Neglecting contributions of $\mathcal{O}(m_\mu^2/Q^2)$, the total, lowest-order cross sections for the DY channels at the partonic level, along with their dependence on mixing and EFT inputs, are:
\begin{align}
 \hat{\sigma}^{d=4}_{\rm DY}(\overline{u}_R d_L \to  \mu^-_L \nu_4) &=  \frac{g^4 \vert U_{\mu4}\vert^2}{2^7  \pi N_c^2 }\frac{(Q^2-m_4^2)^2 (2Q^2+m_4^2)}{Q^4\left[(Q^2 - m_W^2)^2 + (m_W\Gamma_W)^2\right]}\,, \quad 
  \label{eq:collider_parton_xsec_d4}
 \\
 \hat{\sigma}^{d=6}_{\rm DY}(\overline{u}_R d_L  \to  \mu^-_R \nu_4) &= \left( \frac{\left\vert  \mathcal{C}_{H\nu e}\right\vert^2}{4\vert U_{\mu4}\vert^2} \right ) \times  \hat{\sigma}^{d=4}_{\rm DY}(\overline{u}_R d_L  \to  \mu^-_L \nu_4)\,, \quad \text{where} 
  \label{eq:collider_parton_xsec_d6}
 \\
 \mathcal{C}_{H\nu e} &=  v^2 \left[\overline{C}_{H\nu e}\right]_{4\mu}\,.
 \label{eq:collider_parton_coup}
\end{align}
For the $W\gamma$ channel, analogous expressions exist  as does the scaling relationship between $d=4$ and $d=6$ contributions. The fact that phase-space-integrated expressions at $d=4$ and $d=6$ differ by only prefactors is a consequence that both interactions are maximally parity violating: after integration over the $\mu^-$'s polar angle $\theta_\mu$, differences of the form $(1\mp\cos\theta_\mu)$ vanish. 

Importantly, due to  strong kinematic suppression in the DY channel that scales as $\hat{\sigma}_{\rm DY}\sim 1/Q^2$, heavy neutrinos with masses above $m_W$ are produced very close to the kinematic threshold~\cite{Pascoli:2018heg}. To illustrate this nontrivial fact, we plot in Fig.~\ref{fig:issVeto_DYX_LHCMulti_XLO_Mmuv4} the normalized, LO invariant mass distribution of the $(\mu\nu_4)$-system, when mediated by the DY process at $d=4$, for (L) $m_4=150\GeV$ and (R) $m_4=450\GeV$, for $\sqrt{s}=14$, 27 , and 100 TeV. For $m_4=150~(450)\GeV$, we find that the maxima characteristically occur at about $Q_{\rm max}\sim 210~(630)\GeV$, or $Q_{\rm max}\sim1.4\times m_4$. Following Ref.~\cite{Pascoli:2018heg}, we can qualitatively deduce this behavior by noting that the differential cross section with respect to $Q^2$ at the hadronic level can be written for either dimension $d$ as
\begin{align}
 \frac{d\sigma_{\rm DY}}{dQ^2} &= \frac{1}{s} ~ \sum_{ij}\int_{Q^2/s}^1\frac{d\xi_1}{\xi_2}\left[f_i(\xi_1)f_j(\xi_2) + (1\leftrightarrow2)\right] ~ \hat{\sigma}_{\rm DY}, \quad \xi_2 = \frac{Q^2}{s\xi_1}
 \\
 &\equiv \frac{1}{s} \times \sum_{ij} \Phi_{ij}\left(s, Q^2\right) \times \hat{\sigma}_{\rm DY}.
\end{align}
In the first line, the summation runs over all partons that contribute to $pp \to \mu^\pm\nu_4+X$ process at LO via the DY mechanism; $f_i$ and $f_j$ are the usual PDFs that depend on momentum fractions $\xi_1$ and $\xi_2$; and in the second line we expressed the integral over PDFs in terms of the parton luminosities $\Phi_{ij}$. Assuming that $\Phi_{ij}$  are static over small variations of $Q^2$, then the normalized distribution with respect to $Q$,  near the kinematic threshold $Q\sim m_4$, is proportional to the parton-level cross section itself, i.e., 
$(1/\sigma)(d\sigma_{\rm DY}/dQ) \propto Q \times \hat{\sigma}_{\rm DY}$. We can then find the local maximum by taking the first derivative and solving for the root. Assuming that $m_W$ is negligible, we find that the invariant mass of the $(\mu\nu_4)$-system peaks at $Q_{\rm max}=m_4\sqrt{1+\sqrt{3}}\approx 1.65\times m_4$. For nonzero $m_W$, we find a slightly improved estimation of $Q_{\rm max}/m_4 \approx 1.54~(1.64)$ for $m_4 = 150~(450)\GeV$. In reality, parton luminosities quickly fall for even small increases of $Q^2$. This suggests that the true maximum occurs as at lower $Q_{\rm max}$, in agreement with Fig.~\ref{fig:issVeto_DYX_LHCMulti_XLO_Mmuv4}, and ensures the validity of our EFT for $\nu_4$ masses satisfying $m_W \lesssim m_4 \ll \Lambda\sim1\TeV$ at the LHC.

\begin{figure}[!t]
\center 
\includegraphics[width=0.47\textwidth]{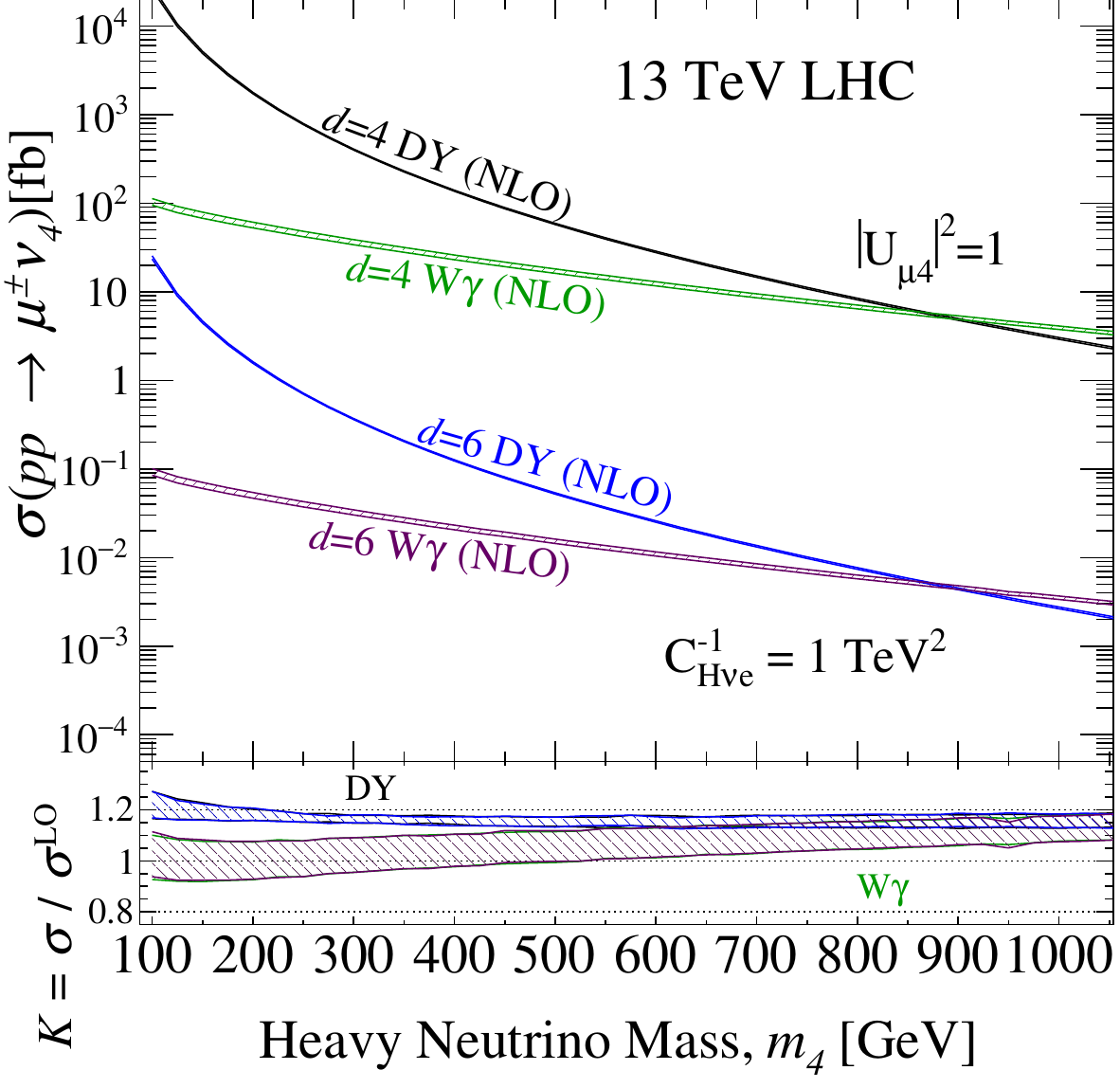}
\includegraphics[width=0.47\textwidth]{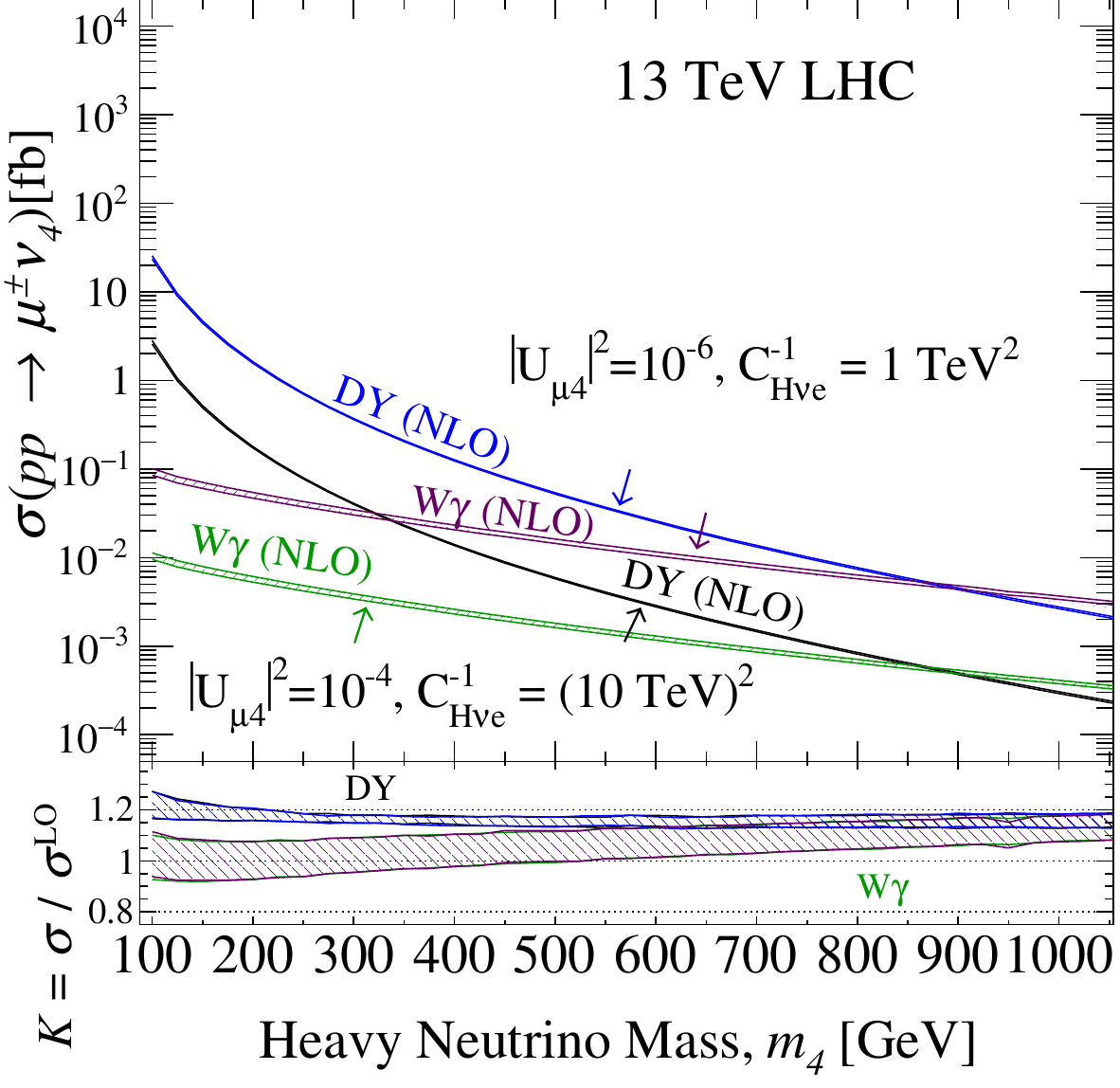}
\caption{
(L) Upper: As a function of $\nu_4$'s mass, the total $pp\to \mu^\pm\nu_4+X$ cross section at NLO in QCD with scale uncertainty (band thickness) at the $\sqrt{s}=13$ TeV LHC for the Drell-Yan (DY) and $W\gamma$ fusion channels purely at dimension $d=4$ assuming $\vert U_{\mu4}\vert^2=1$ (upper curves), and purely at $d=6$ assuming $\left[\overline{C}_{H\nu e}\right]_{4\mu}=(1~{\rm TeV})^{-2}$ (lower curves). Lower: the QCD $K$-factor.
(R) Same as (L) but for the combined $d=4+6$ rate 
assuming (upper curves) $\vert U_{\mu4}\vert^2=10^{-6}$ with $ \left[\overline{C}_{H\nu e}\right]_{4\mu}=(1~{\rm TeV})^{-2}$ and (lower curves)  $\vert U_{\mu4}\vert^2=10^{-4}$ with $\left[\overline{C}_{H\nu e}\right]_{4\mu}=(10~{\rm TeV})^{-2}$.}
\label{fig:lhcxsec}
\end{figure}

The (effective) non-interference of $d=4$ and $d=6$ contributions subsequently allows us to express the total of the two as their linear combination, with coefficients set to the mixing and EFT inputs respectively. For the DY and $W\gamma$ channels, this is given  by
\begin{align}
 \hat{\sigma}^{d=4+d=6}_{{\rm DY}~(W\gamma)} & ~=~ \vert U_{\mu4}\vert^2 \times \hat{\sigma}^{d=4}_{{\rm DY}~(W\gamma)}\left[\vert U_{\mu4}\vert^2=1\right]
 \nonumber\\
 &   \qquad ~+~
\left\vert  \left[\overline{C}_{H\nu e}\right] \right\vert^2 \times (1\TeV)^4 \times
 \hat{\sigma}^{d=6}_{{\rm DY}~(W\gamma)}\left[\left[\overline{C}_{H\nu e}\right]_{4\mu}=(1\TeV)^{-2}\right]
\label{eq:collider_lin_combo}
\\
&~=~ 
 \left( \vert U_{\mu4}\vert^2 + \frac{1}{4}\vert  \mathcal{C}_{H\nu e}\vert^2\right) \times  \hat{\sigma}^{d=4}_{{\rm DY}~(W\gamma)}\left[\vert U_{\mu4}\vert^2=1\right],
 \label{eq:collider_lin_fact}
\end{align}
In the first line,  the total partonic cross section $ \hat{\sigma}^{d=4+6}$ is expressed in terms of $d=4$ and $d=6$ cross sections evaluated that are respectively evaluated assuming
\begin{equation}
\vert U_{\mu4}\vert^2=1 \quad\text{and}\quad \left[ \overline{C}_{H\nu e}\right]_{4\mu}= (1\TeV)^{-2}\,,
\end{equation}
and appropriately re-scaled for arbitrary active-sterile mixing and EFT parameters. In Eq.~\eqref{eq:collider_lin_fact}, we alternatively cast the total cross section solely in terms of the $d=4$ contribution. Regardless of the parametrization, the dependence on mixing and EFT parameters are not altered by convolutions with QCD parton distribution functions (PDFs) nor QCD corrections~\cite{Ruiz:2015zca}. The dependence therefore also hold for cross sections at the hadron level.

With this guidance, we model the production of $\mu^\pm\nu_4$ pairs in LHC collisions by expanding\footnote{
The \texttt{SM\_HeavyN\_vSMEFTdim6\_NLO} and {SM\_HeavyN\_vSMEFTdim6\_XLO} UFOs are available from the URL \href{https://feynrules.irmp.ucl.ac.be/wiki/HeavyN}{feynrules.irmp.ucl.ac.be/wiki/HeavyN}. In the model files, the relevant Wilson coefficient \texttt{CmuN1} is dimensionless and the EFT scale is given by \texttt{Lambda}. These are related to $\overline{C}_{H\nu e}$ by $\left[\overline{C}_{H\nu e}\right]_{4\mu}= \texttt{CmuN1}/\texttt{Lambda}^2$.} the \texttt{HeavyN} UFO libraries~\cite{Alva:2014gxa,Degrande:2016aje} by the operator  $\mathcal{O}_{H\nu e}$. Using the procedure of Refs.~\cite{Degrande:2014vpa,Degrande:2016aje} we can compute the scattering rates for the DY and $W\gamma$ channels up to NLO in QCD at both $d=4$ and $d=6$. For cross section computations, we use the Monte Carlo suite \texttt{MadGraph5\_aMC@NLO}~\cite{Stelzer:1994ta,Alwall:2014hca} and tune our simulation inputs to match the heavy neutrino analysis of Ref.~\cite{Pascoli:2018heg}.

\begin{table}
\small \centering
\begin{tabular}{|c|c|c|c|c|c|}
\hline\hline
$m_4$ & 95\GeV  &   100\GeV &	130\GeV &	150\GeV &	200\GeV\\
$\vert U_{\mu4}\vert^2+  \frac{1}{4}\vert  \mathcal{C}_{H\nu e}\vert^2$ & $3.46\times10^{-3}$  &  
$4.32\times10^{-3}$     & $7.17\times10^{-3}$  &  
$7.97\times10^{-3}$     & $8.88\times10^{-3}$  \\
\hline
$m_4$ & 400\GeV &	600\GeV &	800\GeV &	1000\GeV &	1200\GeV\\
$\vert U_{\mu4}\vert^2+  \frac{1}{4}\vert  \mathcal{C}_{H\nu e}\vert^2$ & $3.01\times10^{-2}$  &
$8.35\times10^{-2}$     &  $2.06\times10^{-1}$  &
$4.41\times10^{-1}$     &  $8.48\times10^{-1}$  \\
\hline\hline
\end{tabular}
\caption{Limits on $d=6$ coupling $\mathcal{C}_{H\nu e} =  v^2 \left[\overline{C}_{H\nu e}\right]_{4\mu }$ and active-sterile neutrino mixing $\vert U_{\mu4}\vert$ as derived from CMS searches for heavy neutrinos with $\mathcal{L}\approx36\invfb$ at $\sqrt{s}=13\TeV$~\cite{Sirunyan:2018mtv}.}
\label{tab2}
\end{table}

In the top-left panel of Fig.~\ref{fig:lhcxsec}, we plot as a function of $m_4$ the total, hadronic $pp\to \mu^\pm\nu_4+X$ cross section for the DY and $W\gamma$ channels, at $d=4$ (upper curves) assuming a benchmark active-sterile mixing of $\vert U_{\mu4}\vert^2=1$, and at $d=6$ (lower curves) assuming a benchmark Wilson coefficient $\left[ \overline{C}_{H\nu e}\right]_{4\mu}= (1\TeV)^{-2}$. The band thickness corresponds to the residual QCD scale uncertainty at NLO and reaches the few percent level for all channels for the range of $m_4$ under consideration. While not shown, PDF uncertainties reach a few percent for all curves. For both sets of scattering processes and for the inputs assumed, we see that the $d=6$ rates are suppressed compared to the $d=4$ rates by a factor $ 4 \vert U_{\mu4}\vert^2 / \vert  \mathcal{C}_{H\nu e}\vert^2  \sim 10^3$ as seen in Eq.~\eqref{eq:collider_parton_xsec_d6}. Consequentially, the $W\gamma$ rate at $d=6$ overtakes its DY counterpart at $m_4\sim900$ GeV, just as it does at $d=4$~\cite{Alva:2014gxa,Degrande:2016aje}. In the lower panel of the same plot we show the QCD factor,
\begin{equation}
K= \sigma^{\rm NLO}/\sigma^{\rm LO}. 
\end{equation}
As anticipated, the impact of QCD corrections to the $d=6$ processes are comparable to the $d=4$ cases. For our choices of SM inputs and QCD evolution scales, we find that QCD corrections at NLO increase the DY $(W\gamma)$ channel at $d=6$ by about $10$-$20\%$~$(0$-$20\%)$.

For the representative combinations of neutrino mixing and EFT parameters:
\begin{align}
\text{Scenario~I~(small mixing/small cutoff)} :&\quad \vert U_{\mu4}\vert^2=10^{-6}, \quad \left[ \overline{C}_{H\nu e}\right]_{4\mu}= (1\TeV)^{-2}\,,\\
\text{Scenario~II~(large mixing/large cutoff)} :&\quad \vert U_{\mu4}\vert^2=10^{-4}, \quad \left[ \overline{C}_{H\nu e}\right]_{4\mu}= (10\TeV)^{-2}\,,
\end{align}
and using the linear relationship of Eq.~\eqref{eq:collider_lin_combo}, we plot in the upper panel of Fig.~\ref{fig:lhcxsec}(R) the total $d=4+6$ contribution to the  $pp\to \mu^\pm\nu_4+X$ production cross section at the $\sqrt{s}=13$ TeV LHC, at NLO in QCD and for the DY and $W\gamma$ channels. For the ``small'' mixing / ``small'' cutoff (upper curves) scenario, which is dominated by physics at $d=6$, we find that hadronic cross section rates at NLO can reach about $\sigma^{d=4+6}\sim10\fb~(5\ab)$ for $m_4\sim150~(1000)$ GeV. Conversely, for the ``large'' mixing / ``large'' cutoff (lower curves), which is dominated by physics at $d=4$, we find rates are about $10\times$ smaller than the first scenario.

\begin{table}
\small \centering
\begin{tabular}{|c|c|c|c|c|c|}
\hline\hline
$m_4$ & 90\GeV  &   100\GeV &	130\GeV &	150\GeV &	200\GeV\\
$\vert U_{e4}\vert^2+  \frac{1}{4}\vert  \mathcal{C}_{H\nu e}\vert^2$ & $6.22\times10^{-3}$  &  
$6.52\times10^{-3}$     & $1.10\times10^{-2}$  &  
$1.43\times10^{-2}$     & $1.36\times10^{-2}$  \\
\hline
$m_4$ & 400\GeV &	600\GeV &	800\GeV &	1000\GeV &	1200\GeV\\
$\vert U_{e4}\vert^2+  \frac{1}{4}\vert  \mathcal{C}_{H\nu e}\vert^2$ & $5.24\times10^{-2}$  &
$1.67\times10^{-1}$     &  $4.28\times10^{-1}$  &
$9.49\times10^{-1}$     &  $1.84$  \\
\hline\hline
\end{tabular}
\caption{Same as Table~\ref{tab2} but for the $e$-flavor channel
and based on search of Ref.~\cite{Sirunyan:2018mtv}.}
\label{tab3}
\end{table}

 At $\sqrt{s}=13$ TeV and with $\mathcal{L}\approx 36\invfb$ of data, searches by the CMS experiment are presently the most constraining for $m_4\approx 100-1000\GeV$ due to their inclusion of the $W\gamma$ channel~\cite{Sirunyan:2018mtv,Sirunyan:2018xiv}, though the sensitivity of ATLAS is projected to be comparable~\cite{Pascoli:2018heg,Fuks:2020att}. In its trilepton analysis~\cite{Sirunyan:2018mtv}, which is sensitive to (pseudo)Dirac and Majorana neutrinos, the CMS experiment reports upper limits on $\vert U_{\mu4}\vert^2$ at the 95\% CL. Using the scaling behavior for the DY and $W\gamma$ channels at $d=4$, limits on active-sterile mixing can be interpreted back as limits on the production cross section of $\mu^\pm\nu_4$ pairs. For a given mass $m_4$, we define the 95\% upper limit on the cross section set by CMS to be
\begin{equation}
 \sigma_{\rm CMS~36\invfb}^{\rm 95\%~CL}(pp\to \mu^\pm\nu_4)  = 
\vert U_{\mu4}^{\rm 95\%~CL}\vert^2 \times \left(
 \sigma^{d=4}_{\rm DY}\left[\vert U_{\mu4}\vert^2=1\right]
 +
  \sigma^{d=4}_{W\gamma}\left[\vert U_{\mu4}\vert^2=1\right]
\right).
\end{equation}
Assuming that the production of $\mu^\pm\nu_4$ pairs is governed by both $d=4$ and $d=6$ operators, then by Eq.~\eqref{eq:collider_lin_fact}, CMS data constrain neutrino mixing and EFT parameters at 95\% CL to satisfy:
\begin{align}
 \left( \vert U_{\mu4}\vert^2 +  \frac{1}{4}\vert  \mathcal{C}_{H\nu e}\vert^2\right) 
 <  
 \frac{\sigma_{\rm CMS~36\invfb}^{\rm 95\%~CL}(pp\to \mu^\pm\nu_4)}
 { \sigma^{d=4}_{\rm DY}\left[\vert U_{\mu4}\vert^2=1\right]
 +
  \sigma^{d=4}_{W\gamma}\left[\vert U_{\mu4}\vert^2=1\right]}.
  \label{eq:cms_limits}
\end{align}
Constraints on mixings and cutoff scales are summarized in Table~\ref{tab2} for representative $\nu_4$ masses. For $m_4\sim 100-1000\GeV$, direct searches by CMS constrain the quantity $(\vert U_{\mu4}\vert^2+  \mathcal{C}_{H\nu e}\vert^2 /4)$ to be below the $3\cdot 10^{-3}-4\cdot 10^{-1}$ level at 95\% CL. In the limit that $\vert U_{\mu4}\vert^2$ vanishes, which can be the case if $\vert U_{\mu4}\vert^2\sim m_1/m_4\lesssim10^{-10}$ as in the high-scale Type I Seesaw, then CMS data only weakly constrain the 
$d=6$ coupling to be $(v^2\left[ \overline{C}_{H\nu e}\right]_{4\mu}) \lesssim 0.13~(1.3)$ for $m_4=100~(1000)$ GeV.

As a brief remark, we note that the above limits assume that the branching fractions (BRs) of $\nu_4$ at dimension $d=4+6$ can be well approximated by those at $d=4$. Due to an interplay between kinematic thresholds and the Goldstone Equivalence Theorem, BR$(\nu_4 \to W\ell)$ at $d=4$ spans about $\sim100\%-50\%$, over the range $m_4\sim90\GeV-1\TeV$. A large $d=6$ contribution could at most increase BR$(\nu_4\to W\mu)$ to 100\%, thereby doubling (at most) the signal yield in the CMS's search for $\nu_4$. Consequentially, the limits reported in Table~\ref{tab2} would be tightened by (at most) a factor of $\sqrt{2}\approx 1.4$. An ambiguity exists, however, because it is possible that other $d=6$ operators  enhance the $\nu_4\to Z\nu_\alpha/H\nu_\alpha$ channels, which then reduce BR$(\nu_4 \to W\mu)$~\cite{Duarte:2016miz}. We avoid this complication by assuming the $d=4$ branching fractions for $\nu_4$.

With the full Run II data set, we anticipate that the sensitivity limit set in Eq.~\eqref{eq:cms_limits} can be improved roughly by a factor of $\sqrt{140\invfb/36\invfb}\sim2$, assuming no improvements to the analysis are made. Likewise, at  the high luminosity LHC (HL-LHC) and again assuming a comparable acceptance and selection rates, one can further improve upon Eq.~\eqref{eq:cms_limits} roughly by a factor of $\sqrt{3\invab/36\invfb}\sim 10$. Combining the independent results of ATLAS and CMS can further improve sensitivity by approximately $\sqrt{2}\sim1.4$.

For the case of electrons (and taus), the entire discussion above holds. In particular, searches for the $pp\to e^\pm \nu_4+X$ process by the CMS experiment constrain $\vert U_{e4}\vert$ and $\vert \overline{C}_{H\nu e}\vert$ to a comparable level as reported for the muon flavor channel~\cite{Sirunyan:2018mtv,Sirunyan:2018xiv}. These constraints are summarized in Table~\ref{tab3} for representative $m_4$. Categorically, limits on effective electron couplings are about $1.5-2\times$ weaker than the limits on effective muon couplings. This is due to the larger background rate in the $e^\pm \nu_4$ channel, and is driven by the mis-identification of jets as electrons~\cite{Sirunyan:2018mtv}. Likewise, limits on effective tau lepton couplings are expected to be even weaker due to the smaller $\tau$-tagging efficiencies~\cite{delAguila:2008cj,Pascoli:2018heg}. We estimate that the full Run II and HL-LHC data sets will improve sensitivity to the $e^\pm \nu_4$ channel to the same degree as the muon case.

As a final remark, we note that operator $\mathcal{O}_{H\nu e}$ can also induce new RH contributions to the same-sign $WW$ scattering process $W^\pm W^\pm \to \ell^\pm \ell^\pm$~\cite{Dicus:1991fk,Fuks:2020att}. At dimension four, the $WW$ process is distinguished from the DY and $W\gamma$ channels by the non-resonant propagation of $\nu_4$ and the quadratic dependence on $U_{\ell 4}$ in its matrix element. Despite the mixing suppression in $W^\pm W^\pm$  scattering, the channel does not experience the same phase space-suppression as in DY and $W\gamma$ since $\nu_4$ is never on-shell. It therefore has a larger $m_4$ reach at the LHC than the other processes~\cite{Fuks:2020att}. The pure $d=6$ contribution to $W^\pm W^\pm$ scattering involves a double insertion of $\mathcal{O}_{H\nu e}$, implying a partonic cross section that is related to the pure $d=4$ cross section by
\begin{align}
 \hat{\sigma}^{d=6}(d d' \to  \ell^-_R \ell^-_R u u') 
 &= \left( \frac{\left\vert  \mathcal{C}_{H\nu e}\right\vert}{2\vert U_{\ell 4}\vert} \right )^4 \times 
  \hat{\sigma}^{d=4}(d d' \to  \ell^-_L \ell^-_L u u') .
\end{align}
 Unlike the DY and $W\gamma$, the $WW$ channel also features a non-vanishing interference term that is linear in both  $ U_{\ell  4}$ and $ \overline{C}_{H\nu e}$. While recent investigations have found improved sensitivity to $\vert U_{\ell 4}\vert^2$ using the $WW$ channel at large $m-4$~\cite{Fuks:2020att}, we  estimate that the sensitivity to $\vert  \mathcal{C}_{H\nu e}\vert$ for $m_4\lesssim1\TeV$ does not reach the levels needed to probe $\Dt a_{\ell}$.

\section{The preferred parameter space}\label{eq:results}

We now discuss the parameter space that can account for the discrepancies in the muon's and electron's anomalous magnetic moment. We begin with the muonic case in Sect.~\ref{sec:results_mu}, where the deviation from the SM is more significant, and address the electronic case in Sect.~\ref{sec:results_el}.

\begin{figure}
\center 
\includegraphics[width=0.47\textwidth]{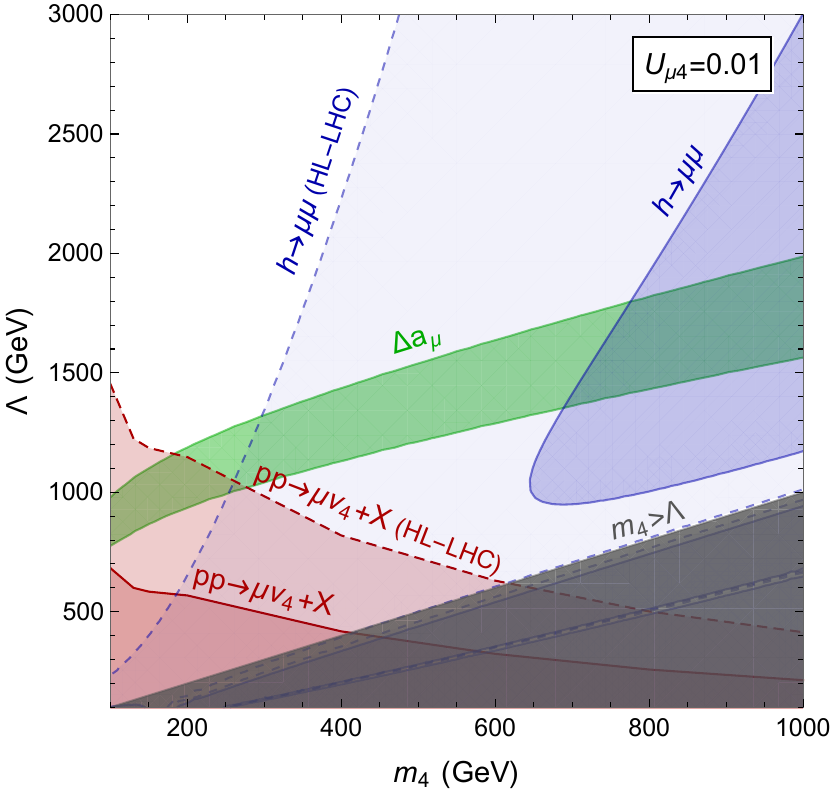}\hfill
\includegraphics[width=0.47\textwidth]{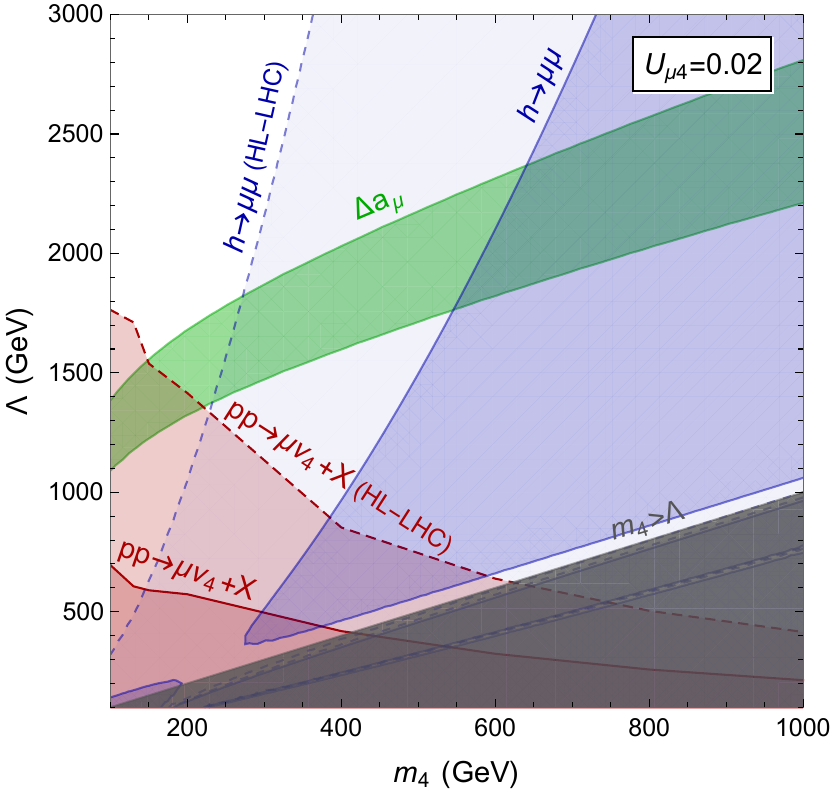}
\caption{The green area depicts parameter space in the $m_4$-$\Lambda$ plane that can account for the $\Delta a_\mu$ discrepancy for a mixing angle of $U_{\mu 4} = 0.01$ (left panel) and $U_{\mu 4} = 0.02$ (right panel). The red region is excluded by CMS and the blue regions are excluded by measurements of the $h\to \mu^+   \mu^-$ branching ratio. The red and blue dashed lines show prospects for the high-luminosity LHC. 
The gray area shows the region with $m_4 \gtrsim \Lambda$, where the $\nu$SMEFT approach breaks down. 
}
\label{plot}
\end{figure} 

\subsection{$\Delta a_\mu$}\label{sec:results_mu}
We present our findings for the $n=1$ case in Fig.~\ref{plot} in the $m_4$-$\Lambda$ plane. In the left (right) panel, motivated by the CKM unitarity bounds we set $\vert U_{\mu4}\vert =10^{-2}$ ($\vert U_{\mu4}\vert =2 \cdot 10^{-2}$), and chose $\left[\bar{C}_{H\nu e}\right]_{4\mu} = -1/\Lambda^2$, where the sign is needed to get a positive $\Delta a_\mu$. The green band corresponds to parameter space that would reconcile the recent {E989} measurement of $\Delta a_\mu$ with the SM prediction of Ref.~\cite{Aoyama:2020ynm}. In the same plot, we depict constraints from several experiments: The red band is the constraint from the CMS experiment on direct $\nu_4$ production as described in {Sect.~\ref{sec:lhcDirect}}.
At present, this does not reach the preferred parameter space yet. The large increase in data from the HL-LHC however is expected to cover the small-$m_4$ part of the green band. The blue area denotes parameter space that is excluded by the observed $h\to \mu^+   \mu^-$ signal strength (the small gap in this area arises from parameter space where dimension-six corrections vanish). The prospects for measurements at the HL-LHC of the $h\to \mu^+   \mu^-$ signal strength are at the $15\%$ level for ATLAS and  $13\%$ for CMS \cite{Cepeda:2019klc}. A combination should give a sensitivity at the $10\%$ level, which is indicated by the dashed blue line in Fig.~\ref{plot}, and covers much of the preferred parameter space. Finally, the dark gray area indicates parameter space where $m_4 > \Lambda$, which is inconsistent with the $\nu$SMEFT framework. The difference between the left and right panel illustrates the sensitivity to $U_{\mu4}$. Larger mixing angles are consistent with larger values of $\Lambda$, but experimental constraints become tighter as well.

\begin{figure}
\center 
\includegraphics[width=0.47\textwidth]{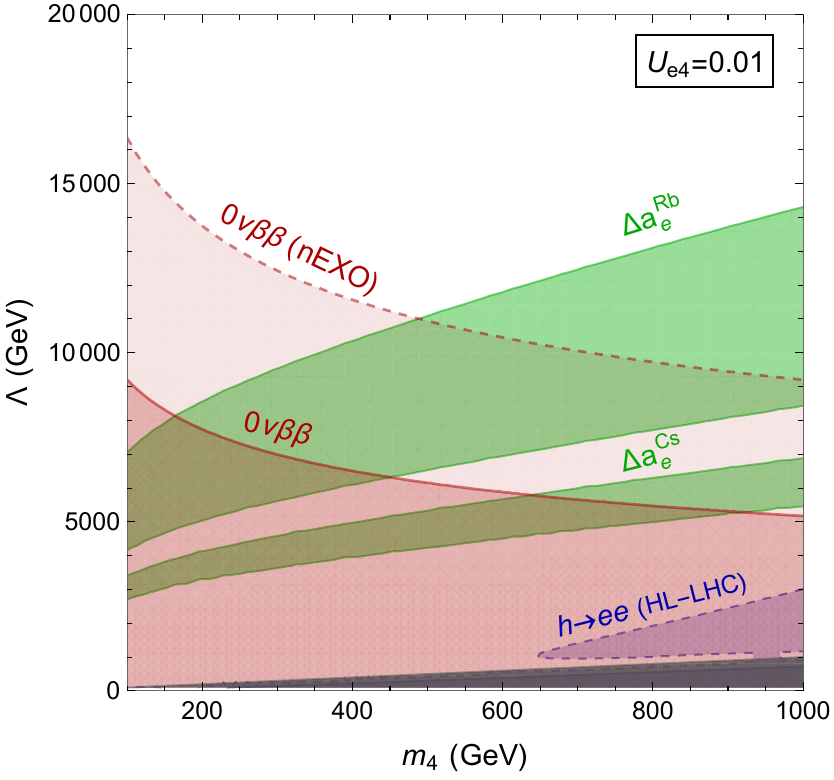}\hfill
\includegraphics[width=0.47\textwidth]{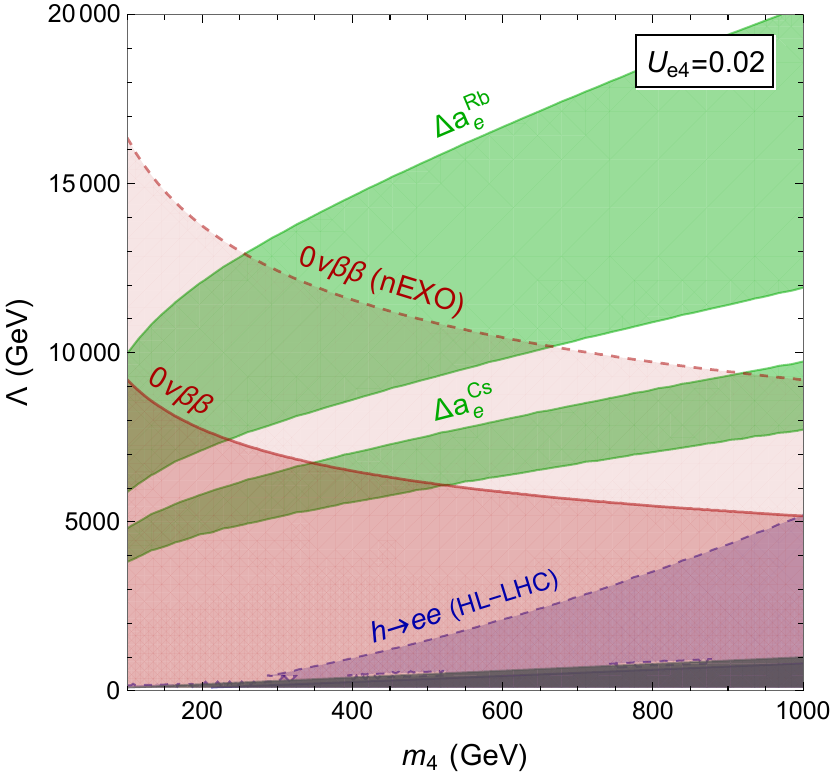}
\caption{
The upper (lower) green area depicts parameter space in the $m_4$-$\Lambda$ plane that can account for the $\Delta a_e^{\rm Rb}$ ($\Delta a_e^{\rm Cs}$) discrepancy for a mixing angle $U_{e 4} = 0.01$ (left panel) and $U_{e 4} = 0.02$ (right panel). The red region depicts the excluded area due to $0\nu\bt\bt$ and the blue regions are excluded by measurements of the $h\to ee$ branching ratio. The red and blue dashed lines show prospects for future $0\nu\bt\bt$ experiments and the high-luminosity LHC. 
The gray area shows the region with $m_4 \gtrsim \Lambda$, where the $\nu$SMEFT approach breaks down. 
}
\label{plote}
\end{figure}

For the representative values of $\vert U_{\mu4}\vert$ that we use, the parameter space that is not ruled out corresponds to a mass range for the sterile neutrino of $m_W < m_4 < 700\,\mathrm{GeV}$, where $0.1 < m_4/\Lambda < 0.5$. Larger masses are ruled out by $h\rightarrow \mu \mu$ measurements, while for lighter masses the constraints on the mixing angle $U_{\mu 4}$ are significantly more stringent such that CMS and ATLAS rule out this part of the parameter space. The green band can be shifted (slightly) upwards by choosing a larger mixing angle closer to the current upper bound $\vert U_{\mu 4}\vert = 0.02$, or downwards by choosing a smaller mixing angle. We find that assuming a smaller mixing angle puts more of the preferred $\Delta a_\mu$ window in reach of searches for $\mu\nu_4$ pairs by ATLAS and CMS.

The plot and discussions in Sect.~\ref{sec:constraints}
make it clear that the $\nu$SMEFT parameter space is very constrained and only a small window is left to account for the muon $g-2$ anomaly. The existing window will be largely covered at Run III of the LHC and the HL-LHC  by searches for direct $\nu_4$ production as well as indirect constraints from $h\to \mu^+   \mu^-$ precision measurements. Additional constraints can arise from low-energy improvements of unitarity tests, which would further shrink allowed values of $U_{\mu 4}$. Finally, following the argument of recent no-lose theorems \cite{Capdevilla:2020qel,Capdevilla:2021rwo}, a confirmation of the $\nu$SMEFT interpretation of $g-2$ results by any of the indirect probes we have discussed suggests that the underlying physics can be directly probed by muon collider proposals that are actively discussed in the literature~\cite{Delahaye:2019omf,Costantini:2020stv,Yin:2020afe,Buttazzo:2020eyl,Ali:2021xlw}.

\subsection{$\Delta a_e$}\label{sec:results_el}

We now turn to $\Delta a_e$. Again considering the case of $n=1$, we show the $m_4 - \Lambda$ plane with $\vert U_{e4}\vert = 0.01$ ($\vert U_{e4}\vert = 0.02$) in the left (right) panel of Fig.\ \ref{plote}. The upper and lower green bands correspond to the regions preferred by $\Dt a_e^{\rm Rb}$ and $\Dt a_e^{\rm Cs}$, which correspond to a different choice of sign, $\left[\bar C_{H\nu e}\right]_{4e} = \pm \frac{1}{\Lambda^2}$, respectively\footnote{This choice of sign hardly impacts the remaining constraints depicted in the figure as they approximately probe $\vert\bar C_{H\nu e}\vert$, given the experimental sensitivities.}. The blue regions are excluded by the current LHC (solid) and prospective HL-LHC (dashed) measurements of Br$(h\to ee)$ discussed in Sect.\ \ref{sec:hll}. The solid blue line falls almost completely falls in the gray area where $m_4\geq\Lambda$, indicating that present $h\rightarrow e^+ e^-$ searches are not accurate enough to set limits. 
The searches for $e\nu_4$ production, discussed in Sect.\ \ref{sec:lhcDirect}, lead to similar limits as in the $\mu$ case. As $\Dt a_e$ can be explained with larger values of $\Lambda$ than $\Dt a_\mu$, these direct limits do not significantly constrain the $\Dt a_e$-preferred regions and are not shown. More stringent constraints come from $0\nu\bt\bt$, see Sect.\ \ref{sec:0nubb}, where the KamLAND-Zen measurement \cite{KamLAND-Zen:2016pfg} currently rules out the solid red region, while the reach of next-generation experiments is shown by the dashed line.

From the left panel, we see that explanations of $\Dt a_e^{\rm Rb}$ require $m_4\gtrsim 200$ GeV and $\Lambda\gtrsim 6.5$ TeV, while  larger $m_4\gtrsim 650$ GeV and smaller $\Lambda >5.5$ TeV are preferred when using $\Dt a_e^{\rm Cs}$. Unlike the case of the muon's $g-2$, the sensitivity of the HL-LHC to $h\to e^+e^-$ is not expected to rule out a significant part of the parameters space. On the other hand, future $0\nu\bt\bt$ experiments can be seen to push up the lower limits to $m_4\gtrsim 500$ GeV and $\Lambda\gtrsim 9$ TeV for explanations of $\Dt a_e^{\rm Rb}$, while they would probe all of the parameter space up to $m_4 = 1$ TeV in the case of  $\Dt a_e^{\rm Cs}$. Similar conclusions hold for the scenario with $\vert U_{e4}\vert=0.02$.

Finally, we note that the experimental limit on the electron EDM, $|d_e| < 1.1\cdot 10^{-29}e$ cm \cite{Andreev:2018ayy}, provides a stringent constraint on the imaginary part of the combination of couplings, 
\begin{equation}
\left|\sum_{j=4}^{n+3} \frac{m_j}{v}\,w(x_j)\,{\rm Im}\left( U_{e j}   \left[v^2\bar  C_{H\nu e}\right]_{j e}\right) \right|\leq 2.2\cdot 10^{-11}\,.
\end{equation} 
The real part of the same combination induces $\Dt a_e$ and would need to be roughly  six orders of magnitude larger to contribute at the level of $\Dt a_e^{\rm Cs,Rb}$, implying explanations of these discrepancies will need be CP-conserving to good precision. In contrast, the constraint due to $d_\mu$ \cite{Bennett:2008dy} allows the corresponding combination of muon couplings to have an imaginary part that is larger than the real part needed to explain $\Dt a_\mu$.


\subsection{$\Delta a_\mu$ and $\Delta a_e$}\label{amuae}
As the sterile neutrino operator $\mathcal O_{H\nu e}$ seems to be able to explain the discrepancies in both $a_e$ and $a_\mu$, a natural question is whether it could explain both types of anomalies simultaneously. For the case of a single sterile neutrino that we have focused on so far, the severe limits from searches for $\mu \rightarrow e \gamma$  exclude this possibility. In this scenario one can express the branching ratio as $\textrm{BR}(\mu \rightarrow e \gamma)  \sim \big|\frac{\Dt a_e}{m_e\xi}\big|^2 +\big|\xi \frac{\Dt a_\mu}{m_\mu}\big|^2 $, with $\xi = U_{e4}/U_{\mu4}$.  Using $\Dt a_e^{\rm Cs}$, this leads to a lower bound on the prediction for the branching fraction of 
\begin{equation}
\textrm{BR}(\mu \rightarrow e \gamma)_{\rm min}  = \tau_\mu \frac{\al_{em} m_\mu^3}{8}\Bigg|\frac{\Dt a_e\Dt a_\mu}{m_e m_\mu}\Bigg| \simeq  1.4\cdot 10^{-4}\,.
\end{equation}

Such a value is clearly well above the current limit of $\textrm{BR}(\mu \rightarrow e \gamma) < 4.2 \cdot 10^{-13}$ \cite{TheMEG:2016wtm}. These limits can, in principle, be  avoided by introducing two sterile states, i.e., $n=2$, with each mass eigenstate coupling almost exclusively to the electron and muon,  indicating a hierarchy within the active-sterile mixing sector. In such scenarios, the $\mu\to e\g$ limits would require certain mixing matrix elements, e.g.\ $U_{e 5}$ and $U_{\mu 4}$ for $n=2$, to be tuned to zero to very high precision compared to the nonzero elements, $U_{e4}$ and $U_{\mu 5}$.

\section{Conclusions}\label{conclusion}

In this work we investigated contributions to the anomalous magnetic moments of leptons in the framework of the neutrino-extended Standard Model Effective Field Theory. We investigated which $\nu$SMEFT operators can give sizable contributions to $\Delta a_e$ and $\Delta a_\mu$ and concluded that remarkably only one dimension-six $\nu$SMEFT operator is relevant for $\Lambda \gg v$. This operator effectively induces a right-handed, leptonic charged current, which modifies leptonic anomalous magnetic moments at one loop. Due to a chiral enhancement and experimental constraints the contribution can be sizable if the associated sterile neutrino has a mass above $m_W$. 

These results extend the contributions of SMEFT operators obtained in Ref.\  \cite{Aebischer:2021uvt} to the $\nu$SMEFT. Following  the notation of Ref.~\cite{Aebischer:2021uvt} and assuming BSM physics arises at a scale $\Lambda = 1$ TeV, we can write the complete $\nu$SMEFT contribution to $\Delta a_\mu$ and $\Delta a_e$ as
\bea
\Dt a_\ell &=& \frac{m_\ell}{m_\mu}{\rm Re}\Bigg[ {1.8_\mu\atop 1.7_e}\cdot 10^{-4}\widetilde C_{\substack{eB\\\ell\ell}}-{9.6_\mu \atop 9.3_e}\cdot 10^{-5}\widetilde C_{\substack{eW\\\ell\ell}}-{10_\mu\atop 9.7_e}\cdot 10^{-6}\widetilde C_{\substack{lequ\\ \ell\ell33}}^{(3)}\nn\\
&&-(1.9+0.2 c_T^{(c)})\cdot 10^{-7}\widetilde C^{(3)}_{\substack{lequ\\ \ell\ell22}}
+1.4\cdot 10^{-8}\widetilde C_{\substack{lequ\\ \ell\ell33}}^{(1)}-4.9\cdot 10^{-9}\widetilde C_{\substack{le\\\ell33\ell}}\nn\\
&&+\left(3.4c_T-0.4\right)\cdot 10^{-9}\widetilde C_{\substack{lequ\\\ell\ell11}}^{(3)}+\left(2.1c_T^{(c)}+{10_\mu\atop 9.7_e}\right)\cdot 10^{-10}\widetilde C_{\substack{lequ\\\ell\ell22}}^{(1)}\nn\\
&&
-2.9\cdot 10^{-10}\widetilde C_{\substack{le\\\ell22\ell}}
+\frac{m_\ell}{m_\mu}\left(2.9\cdot 10^{-9}\widetilde C_{HWB}-2.3\cdot 10^{-9}\widetilde C_{HB}\right)\nn\\
&&-1.1\cdot 10^{-7}\sum_{j=4}^{3+n} \frac{m_j}{m_W} w(x_j)U_{\ell j}\left[\widetilde C_{H\nu e}\right]_{j\ell }
\Bigg]\,.
\eea
Here the first four lines are due to SMEFT interactions, where the most significant contributions come from the leptonic dipole moments, $C_{eB}$ and $C_{eW}$, and semi-leptonic four-fermion interactions, $C_{lequ}^{(3)}$. The dimensionless couplings $\widetilde{C}$ are related to the usual SMEFT Wilson coefficients by $\widetilde C = \Lambda^2 C(\Lambda)$ and we dropped terms whose coefficients for $\ell=\mu$ ($\ell=e$) are smaller than $10^{-9}$ ($10^{-12}$).  $c_T$  and $c_T^{(c)}$ are related to non-perturbative contributions which are proportional to the matrix elements of light quarks and charm quarks, respectively, see Ref.\ \cite{Aebischer:2021uvt} for details. Finally, the last line gives the $\nu$SMEFT contribution where we used  $\left[\widetilde C_{H\nu e}\right]_{j\ell} = \Lambda^2\left[\bar  C_{H\nu e}(\Lambda)\right]_{j\ell} $ with $\Lambda = 1$ TeV, where the loop function $w(x)\to 1/2$ in the limit of $x\to \infty$.

Evaluating the contribution due to $\bar C_{H\nu e}$, we find a small window of preferred masses, $100 < m_4/(1\,\mathrm{GeV}) < 700$, for $\vert U_{\mu 4}\vert =\mathcal O(10^{-2})$, and $v/\Lambda = \mathcal O(0.2)$, where $\Delta a_\mu$ can be accounted for. This parameter space will be largely tested by future measurements at the high-luminosity LHC.  Similarly, for $\vert U_{e4}\vert =0.01$ and $m_4>200$ GeV a values of $\Lambda > 6.5$ TeV exists where $\Dt a_e^{\rm Rb}$ can be accommodated. In this case, future $0\nu\bt\bt$ experiments will be able to probe a significant part of the parameters space, up to $m_4<1$ TeV.

The main focus of this work is to identify whether new $\nu$SMEFT contributions could be relevant and not to speculate about potential UV completions. Nevertheless, it is worthwhile to consider what kind of models could reproduce the low-energy EFT. The dimension-six $\mathcal{O}_{H\nu e}$ operator could hint towards a RH $W_R^\pm$ boson that couples mainly to muons or electrons and mixes with the SM $W^\pm$ fields. Similar $W_R^\pm$ bosons appear for instance in left-right symmetric models. However, these models typically couple the $W_R$ to first-generation quarks as well, in which case $m_{W_R} \sim 1$ TeV, as required in this work to account for $\Delta a_\mu$, is already ruled out by searches for $W_R$ production at the LHC~\cite{Ruiz:2017nip,Aad:2019hjw,Sirunyan:2019vgj}. Models can be envisioned where $W_R^\pm$ does not couple to quarks at all or at least not to first-generation quarks to avoid such constraints. 

Other possible scenarios are models with vector-like leptons which have been studied in the context of explanations of other anomalies such as the first-row CKM unitarity and lepton-flavor universality violation in $B$ meson decays \cite{Megias:2017dzd,Poh:2017tfo, Endo:2020tkb, Crivellin:2020ebi}. These hypothetical leptons have been searched for at the LHC \cite{Aad:2020fzq, Aad:2015dha, Sirunyan:2019ofn}. In such models the vector-like leptons can mix with the RH neutrinos and muons, giving rise to the $\mathcal{O}_{H\nu e}$ operator at tree level. For example, in the case of the $SU(2)$ doublet vector-like lepton $\Delta_1\sim (1,2,-1/2)$, we obtain $C_{H\nu e}\sim \lambda_{\Delta_1\nu_R}\lambda_{\Delta_1\mu_R}/m^2_{\Delta_1}$ with $\lambda_{ij}$ being the Yukawa coupling involving $i$ and $j$ particles. The parameter space for the successful $\Delta a_{\mu}$ explanation can be achieved if $ \lambda_{\Delta_1\nu_R}\lambda_{\Delta_1\mu_R}\sim \Or(1)$ and $m_{\Delta_1}\sim \Or(1)~$TeV.  Current LHC limits are below 1 TeV \cite{Sirunyan:2019ofn}, while future colliders are expected to reach the TeV range \cite{Bhattiprolu:2019vdu}. We leave a more detailed analysis of possible scenarios to future work. 


\bigskip
\section*{Acknowledgements}The work of VC, KF and EM is supported  by the US Department of Energy through the Los Alamos National Laboratory. KF is also supported by the LANL LDRD Program.  Los Alamos National Laboratory is operated by Triad National Security, LLC, for the National Nuclear Security Administration of U.S.\ Department of Energy (Contract No.\ 89233218CNA000001).
WD is supported by  U.S.\ Department of Energy Office of Science, under contract  DE-SC0009919.
RR acknowledges the support of Narodowe Centrum Nauki under Grant No. 2019/34/E/ST2/00186.

\bibliography{bibliography}

\end{document}